\documentclass[sigconf]{acmart}

\usepackage{booktabs}      
\usepackage{siunitx}       
\usepackage{subcaption}    
\usepackage{multirow}      
\usepackage{makecell}      
\usepackage{pifont}        
\usepackage{xcolor} 
\usepackage{mdframed}
\NewDocumentEnvironment{tcolorbox}{O{}}{%
  \begin{mdframed}[
    backgroundcolor=black!5,
    linecolor=black!25,
    linewidth=0.1pt,
    roundcorner=5pt,
    skipabove=0pt,
    skipbelow=0pt,
    innertopmargin=3pt,
    innerbottommargin=3pt,
    innerleftmargin=5pt,
    innerrightmargin=5pt,
    nobreak=true
  ]
}{%
  \end{mdframed}
}

\usepackage{algorithm}
\usepackage{algpseudocode}

\usepackage{enumitem}

\usepackage{xspace}

\newcommand{\mymodel}{\textsf{NES}\xspace}
\newcommand{\mytab}{\texttt{Tab}\xspace}
\newcommand{\mycode}[1]{\texttt{\footnotesize #1}} 

\newtheorem{definition}{Definition}

\AtBeginDocument{%
  }

\title[NES: An Instruction-Free, Low-Latency Next Edit Suggestion Framework]{NES: An Instruction-Free, Low-Latency Next Edit Suggestion Framework Powered by Learned Historical Editing Trajectories}

\author{Xinfang Chen}
\authornote{Equal contribution.}
\email{cxf453404@antgroup.com}
\affiliation{%
  \institution{Ant Group, China}
\city{}
\country{}
  }

\author{Siyang Xiao}
\authornotemark[1]
\email{xiaosiyang.xsy@antgroup.com}
\affiliation{%
  \institution{Ant Group, China}
  \city{}
\country{}
  }

\author{Xianying Zhu}
\email{zhuxianying.zxy@antgroup.com}
\affiliation{%
  \institution{Ant Group, China}
  \city{}
\country{}
  }

\author{Junhong Xie}
\email{qingyi.xjh@antgroup.com}
\affiliation{%
  \institution{Ant Group, China}
  \city{}
\country{}
  }

\author{Ming Liang}
\email{liangming.liang@antgroup.com}
\affiliation{%
  \institution{Ant Group, China}
  \city{}
\country{}
  }

\author{Dajun Chen}
\email{chendajun.cdj@antgroup.com}
\affiliation{%
  \institution{Ant Group, China}
  \city{}
\country{}
  }

\author{Wei Jiang}
\email{jonny.jw@antgroup.com}
\affiliation{%
  \institution{Ant Group, China}
  \city{}
\country{}
  }

\author{Yong Li}
\authornote{Corresponding author.}
\email{liyong.liy@antgroup.com}
\affiliation{%
  \institution{Ant Group, China}
  \city{}
\country{}
  }
  \author{Peng Di}
  \authornotemark[2]
\email{dipeng.dp@antgroup.com}
\affiliation{%
  \institution{Ant Group, China \& UNSW Sydney}
  \city{}
\country{}
  }

\renewcommand{\shortauthors}{Chen et al.}

\copyrightyear{2026}
\acmYear{2026}
\setcopyright{cc}
\setcctype{by}
\acmConference[FSE Companion '26]{34th ACM Joint European Software Engineering Conference and Symposium on the Foundations of Software Engineering}{July 05--09, 2026}{Montreal, QC, Canada}
\acmBooktitle{34th ACM Joint European Software Engineering Conference and Symposium on the Foundations of Software Engineering (FSE Companion '26), July 05--09, 2026, Montreal, QC, Canada}
\acmDOI{10.1145/3803437.3805244}
\acmISBN{979-8-4007-2636-1/2026/07}

\begin{document}

\begin{abstract}

Code editing is a frequent yet cognitively demanding task in software development. Existing AI-powered tools often disrupt developer flow by requiring explicit natural language instructions and suffer from high latency, limiting real-world usability. We present \mymodel (\textbf{N}ext \textbf{E}dit \textbf{S}uggestion), an instruction-free, low-latency code editing framework that leverages learned historical editing trajectories to implicitly capture developers’ goals and coding habits. \mymodel features a dual-model architecture: one model predicts the next edit location, and the other generates the precise code change—both without any user instruction. Trained on our open-sourced \emph{SFT} and \emph{DAPO} datasets~\footref{footdataset}, \mymodel achieves state-of-the-art performance (75.6\% location accuracy, 27.7\% exact match rate) while delivering suggestions in under 250ms. Deployed at Ant Group, \mymodel serves over 20,000 developers through a seamless Tab-key interaction, achieving effective acceptance rates of 51.55\% for location predictions and 43.44\% for edits—demonstrating its practical impact in real-world development workflows. 

\end{abstract}



\maketitle

\section{Introduction}
\label{sec:introduction}

With Large Language Models (LLMs) demonstrating remarkable capabilities in various aspects of software development~\cite{Di2023CodeFuse13BAP, qwen3technicalreport, Wang2024RLCoderRL, Guo2024DeepSeekCoderWT, REPOFUSE, Rozire2023CodeLO, Ding2024VulnerabilityDW,codefuse2025samplemattersleveragingmixtureofexperts}, AI-powered development tools such as GitHub Copilot~\cite{copilot}, Cursor~\cite{cursor} and Zeta~\cite{zeta} are revolutionizing modern software development workflows by leveraging LLMs for code editing tasks~\cite{Wang2021CodeT5IU, Zhang2022CoditT5PF, GRACE, wei2024coeditor, Liu2024CoEdPilotRC}. Empirical studies further highlight that code editing, which includes activities such as modifying~\cite{Li2023InstructCoderIT, guo2025codeeditorbenchevaluatingcodeediting}, refactoring, and maintaining existing code, is one of the most frequent and critical activities throughout the lifecycle of a software project~\cite{6693078}. In particular, code completion task can be formally regarded as a restricted instance of code editing, constrained to insertions immediately following the cursor position within a partially written syntactic construct (e.g., statement, expression, or function body). The importance of efficient and accurate code editing cannot be overstated, as it directly impacts the productivity and quality of software development.



Existing research efforts
have made significant advancements, particularly in translating explicit natural language instructions into code edits. However, these approaches face critical \emph{limitations} that hinder their effective integration into a developer's workflow.
\emph{On one hand}, most existing methods, such as CodeT5~\cite{Wang2021CodeT5IU}, CoditT5~\cite{Zhang2022CoditT5PF},  Coeditor~\cite{wei2024coeditor}, 
CoEdPilot~\cite{Liu2024CoEdPilotRC}, EfficientEdit~\cite{wang2025reusegenerateacceleratingcode} and so on,
\textbf{rely heavily on human instruction input} to maintain a coherent reference chain, disrupting programming flow and increasing cognitive load. Additionally, tools that fail to capture users' habitual coding behavior often generate suggestions misaligned with their natural coding style.
\emph{On the other hand}, code editing tasks are inherently time-sensitive, with users expecting feedback within 1s\cite{card2018psychology}. But existing tools~\cite{qwen3technicalreport,kimiteam2025kimik15scalingreinforcement,codefuse2025samplemattersleveragingmixtureofexperts,TheC3,Guo2024DeepSeekCoderWT} rely on computationally intensive processes (e.g., search, text understanding, reasoning), creating a tradeoff between low-latency and precise suggestions. This \textbf{significant latency} undermines real-time interaction and degrades user experience. 
Overcoming these limitations can undoubtedly enhance code editing in real-world development. However, it necessitates modifying the method of acquiring cues, which poses significant challenges in accurately identifying the editing location and determining the content within the desired timeframe.


Our approach is based on an empirical observation: users' habitual behaviors and coding objectives are often embedded within their historical editing patterns (e.g., repetitive refactoring actions or managing cross-file dependencies), which provide valuable contextual cues. Using historical editing patterns instead of relying on human instructions can minimize interruptions to the programming flow, maintain user-style consistency in suggestions, and enhance both the accuracy and efficiency of reference generation.

We propose \mymodel (\textbf{N}ext \textbf{E}dit \textbf{S}uggestion), an LLM-driven code editing framework designed for an \textbf{instruction-free} and \textbf{low-latency (feel like zero-latency)} experience. Comprising a synergistic dual-model architecture trained on the proposed \emph{Supervised Fine-Tuning (SFT)}~\cite{wei2022finetuned} and \emph{Dynamic sAmpling Policy Optimization (DAPO)}~\cite{yu2025dapoopensourcellmreinforcement} datasets, \mymodel adapts to user habits, integrates seamlessly into workflows, and enhances productivity by implicitly learning developer intent.
\begin{itemize}[leftmargin=*]
    \item \textbf{\mymodel-Location Model} utilizes the developer's historical editing patterns to predict the most probable next editing location, facilitating proactive and efficient navigation across the codebase. 
    \item \textbf{\mymodel-Edit Model} analyzes the developer's editing history, including refactoring patterns and API call sequences, to generate personalized and precise code modifications for the current edit. 
\end{itemize}
To improve user experience, \mymodel leverages Prefix Caching (PC)~\cite{zheng2024sglangefficientexecutionstructured} and Speculative Decoding (SD)~\cite{Speculative,Chen2023AcceleratingLL} techniques to achieve substantial inference speedups.

Proven at scale with over 20,000 developers at Ant Group, \mymodel is an industry-ready solution that streamlines complex coding tasks. Its core innovation is a continuous workflow where developers use simple \mytab key sequences (\mytab→\mytab→\mytab) to perform operations like refactoring. This unique interaction model eliminates manual steps, significantly reducing cognitive load and context-switching to create a highly efficient and immersive coding experience.



We demonstrate the practical effectiveness of \mymodel using both open-source and industrial datasets. Our model achieves SOTA performance, with the Location Model reaching 75.6\% accuracy in predicting edit placement and the Edit Model delivering a 27.7\% Exact Match Rate (EMR) for generating intent-aligned code. By optimizing for Nvidia L20 GPUs, we ensure efficiency with an inference time of under 250ms. Additionally, our proposed SFT and DAPO datasets significantly improve existing Qwen2.5-Coder-7B and Seed-Coder-8B-Instruct models.


Our key contributions are as follows.

\begin{itemize}[leftmargin=*]
    
    

\item We introduce \mymodel\footnote{Unless otherwise specified, \mymodel refers to the Qwen3-4B+SFT+DAPO model. See Section~\ref{subsec:industrial_tools} for further details.}, a dual-model, LLM-powered code editing framework that leverages historical code edit patterns. Operating in an instruction-free, low-latency setting, it eliminates the need for explicit natural language input from users.

\item We construct and open-source high-quality \emph{SFT} and \emph{DAPO} datasets\footnote{A portion of the dataset is publicly available at \url{https://huggingface.co/datasets/codefuse-ai/CodeFuse\_codeedit}. The remaining data will be released incrementally.\label{footdataset}}, which significantly boost performance. These datasets have played a critical role in achieving dramatic improvements—by orders of magnitude—in accuracy and similarity metrics across various code editing benchmarks.

\item We have integrated and deployed \mymodel across Nvidia L20 GPUs within Ant Group, serving 20,000+ developers. The implementation of various inference optimizations ensures that \mymodel delivers real-time responses (under 250ms), culminating in a more fluid user experience with effective acceptance rates of 43.44\% for code edits and 51.55\% for location prediction.

\end{itemize}


\section{Motivating Example}
\label{sec:example}

\begin{figure}[t] 
    \centering
    \includegraphics[width=1.0\linewidth]{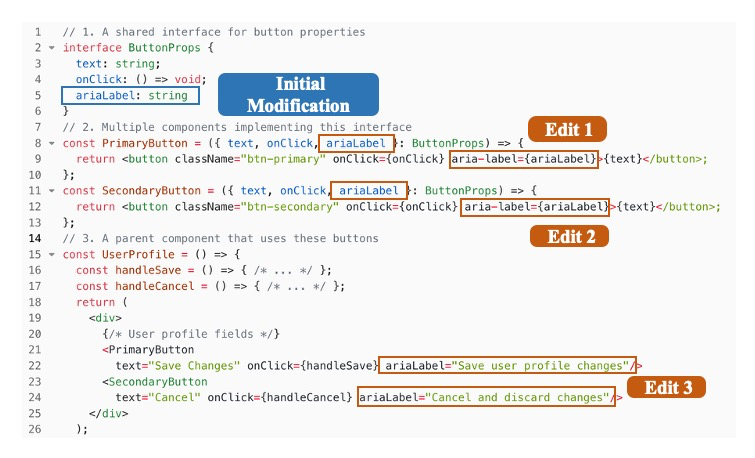}
    \vspace{-4ex}
    \caption{A motivating example of \mymodel. \label{example}}
\vspace{3ex}
\includegraphics[ width=1.0\linewidth]{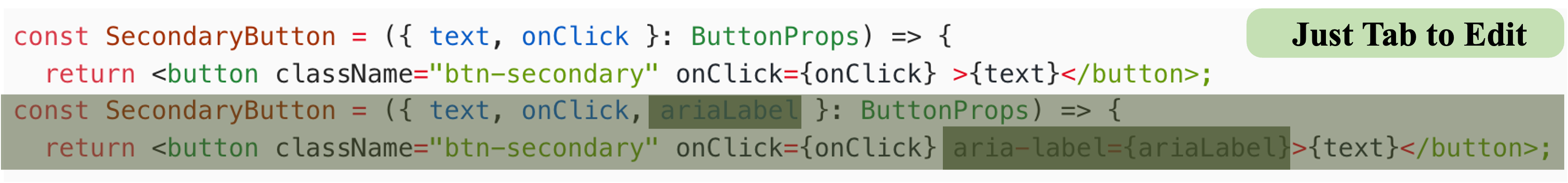}
\vspace{-4ex}
    \caption{\mymodel's suggestion accepted by the developer with a single keystroke.\label{nes_result}}
    \vspace{-2ex}
\end{figure}

This section motivates \mymodel through a representative refactoring task in modern web development. For example, a software engineer enhancing UI components must add an \mycode{aria-label} property to all interactive buttons for improved screen reader accessibility (see Figure~\ref{example} for a common design pattern with a shared interface and multiple implementing components).


Without an adaptive assistance framework like \mymodel, the developer must undertake a series of laborious steps:
\begin{enumerate}[leftmargin=*]
    \item \textbf{Initial Modification:} The developer updates the \mycode{ButtonProps} interface to include \mycode{aria-label:string;} property (Line 5).
    \item \textbf{Edit 1:} The \mycode{PrimaryButton} is manually updated to de-structure the new prop and apply it to the \mycode{<button>} element (Line 8-9).
    \item \textbf{Edit 2:} The same changes are repeated for the \mycode{SecondaryButton} component(Line 11-12).
    \item \textbf{Edit 3:} The developer traces component usages, such as in the \mycode{UserProfile} component, and adds \mycode{aria-label} values to instances of  \mycode{<PrimaryButton/>}(Line 22) and \mycode{<SecondaryButton/>}(Line 24).
    
\end{enumerate}


The process involves manual navigation, repetitive edits, and significant cognitive effort to track changes. Although LLM-driven tools can assist with locating and making edits, instruction-based tools, such as CodeT5~\cite{Wang2021CodeT5IU}, CoditT5~\cite{Zhang2022CoditT5PF}, Coeditor~\cite{wei2024coeditor}, and CoEdPilot~\cite{Liu2024CoEdPilotRC}, disrupt the developer's workflow by requiring precise commands and waiting for responses, introducing additional cognitive overhead. To address these challenges, an instruction-free tool is essential, which is where \mymodel comes in.

\mymodel streamlines this task by predicting the developer's implicit intent based on their actions, enabling a seamless, fluid interaction.

\begin{enumerate}[leftmargin=*]

\item \textbf{Intent Seeding:} The developer initiates the process by intentionally adding the \mycode{aria-label: string;} property to the \mycode{ButtonProps} interface, signaling a cue for the \mymodel framework.

\item \textbf{Proactive Edit Suggestion (\mymodel-Edit):} When the cursor moves near the \mycode{PrimaryButton} definition, the \mymodel-Edit model detects the dependency change and familiar refactoring pattern, proactively suggesting updates (see Figure~\ref{nes_result}) to the component's signature and implementation.
 
\item \textbf{Predictive Navigation and Chained Editing (\mymodel-Location + \mymodel-Edit):} The \mymodel-Location model predicts the next logical change at the \mycode{SecondaryButton} component, guiding the cursor to its definition. Upon arrival, the \mymodel-Edit model suggests the required refactoring, accepted with a single \mytab press.

\item \textbf{Cascading Assistance:} The framework continues predicting and assisting, leading to call sites in \mycode{UserProfile}. Here, \mymodel-Edit suggests context-specific values (e.g., \mycode{aria-label="Save Changes"}) for the new property, with each step accepted via \mytab.
\end{enumerate}

A tedious, error-prone manual process is streamlined into an efficient "\mytab→\mytab→\mytab" workflow. Powered by high-quality datasets, \mymodel predicts both the location (\mymodel-Location) and content (\mymodel-Edit) of edits without requiring explicit instructions. This reduces cognitive load, minimizes context switching, accelerates inference and delivers an immersive development experience.

\section{Methodology}





\begin{figure*}
    \centering
    \includegraphics[width=0.95\linewidth]{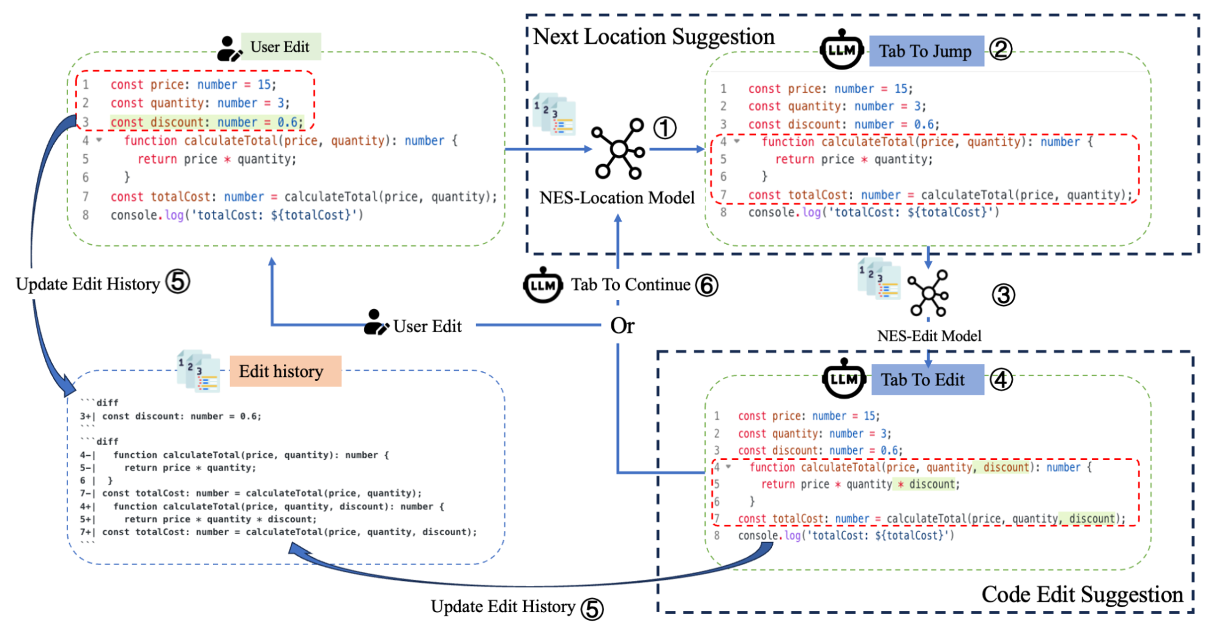}
    \caption{Illustration of the \mymodel framework’s
workflow.}
    \label{fig:overview}
\end{figure*}

\subsection{An Overview of \mymodel Framework}
Figure \ref{fig:overview} provides a detailed illustration of the \mymodel framework's workflow. The framework's functionality is initiated by the developer's initial editing action, subsequently entering a dynamic interactive cycle: 
\begin{itemize}[leftmargin=*]

\item \textbf{Next Location Suggestion:} The \mymodel-Location Model leverages real-time editing trajectories~\textcircled{1} to predict the next likely edit location. The developer can rapidly jump to this position by pressing the \mytab{} key or simply disregard the suggestion~\textcircled{2}.


\item \textbf{Code Edit Suggestion:} Once at the location, the \mymodel-Edit Model generates a code edit by integrating historical patterns with the current code context~\textcircled{3}. This suggestion can be accepted with the \mytab{} key, or the developer can continue editing manually~\textcircled{4}, which in turn updates the edit history~\textcircled{5}.


\item \textbf{Dynamic loop:} Each accepted suggestion triggers the next prediction, creating a continuous assistance loop~\textcircled{6}. This process transforms complex refactoring tasks into a fluid \mytab sequence, significantly boosting developer productivity.
\end{itemize}

\noindent To implement this framework, we focused on four key areas:

\paragraph{Incremental Difference Detection} We developed a real-time code change monitor based on incremental computation. This framework accurately captures each developer editing operation, thereby constructing a comprehensive and detailed historical editing trajectory. These trajectory data serve as crucial inputs for subsequent model training and prediction processes. 

\paragraph{High-Quality Training Dataset Construction} Utilizing the collected authentic historical editing trajectories, we constructed a high-quality training dataset. This dataset encompasses not only sequences of editing operations but also contextual information for each edit. Therefore, this dataset can enable the training of both the \mymodel-Location Model and the \mymodel-Edit Model with nuanced understanding of editing patterns and their surrounding code environments. 

\paragraph{Model Training} To endow \mymodel with the capability for location and code editing suggestion, we designed a two-stage training methodology. Initially, both the \mymodel-Location and \mymodel-Edit models undergo \emph{SFT}~\cite{InstructGPT} on our large-scale historical editing datasets, enabling them to learn the fundamental patterns of code modification. Subsequently, we employ \emph{Decoupled Clip} and \emph{DAPO}~\cite{yu2025dapoopensourcellmreinforcement}, a reinforcement learning technique using high-quality preference data, to further refine the models\footnote{DAPO is a reinforcement learning method that uses dynamic sampling of high-quality preferences to refine the policy, while Decoupled Clip stabilizes training by decoupling the clipping mechanism in policy updates.}.
This second stage aligns their behavior more closely with real-world developer intent and utility, significantly boosting the practical value of \mymodel.

\paragraph{Model Inference Optimization} Acknowledging the constraints of inference resources in practical application environments and the stringent low-latency requirements inherent in real-time editing suggestion, we conducted in-depth optimization of the model inference process, and achieved real-time responsiveness to developer editing behaviors, ensuring the fluidity and practicality of the \mymodel framework.

We present the details of these critical components in the following sections, respectively.

\subsection{Incremental Difference Detection\label{Incremental Difference Detection}}


We developed an efficient incremental difference detector built upon our AI-powered IDE, designed to accurately capture real-world code editing trajectories of developers during real-world programming activities. 
The algorithm operates in two primary steps: incremental difference calculation and difference merging, where "Overlap" denotes intersecting regions of two editing operations.

\subsubsection{Incremental Difference Calculation\label{sec:diffcalc}}


To ensure real-time responsiveness, we replaced conventional file-wide difference calculation methods with an incremental approach. The core principle is to narrow the computational scope from the entire file to only the code segments actively being modified by the developer. 
By focusing solely on code segments actively modified by the user, our AI-powered IDE captures edit regions in real-time and applies a sequence-matching method to these localized segments, minimizing computational overhead and enabling instant change detection.

To provide the model with precise and unambiguous context, we designed the custom NES diff format. 
As illustrated in Table~\ref{tab:simple}, our format enriches the standard diff by prefixing every line—whether added (\texttt{+}), deleted (\texttt{-}), or unchanged—with its absolute line number. 
The NES diff format provides clearer code modification and line number change information for the model, directly improving the accuracy of location and code editing tasks.



\subsubsection{Instant Difference Merge}

The high-frequency differential computation in our framework results in the acquisition of extensive granular editing histories. Utilizing such fragmented data for model training would potentially lead to the generation of incomplete and atomized suggestions, necessitating frequent user interactions to adopt multiple fragmentary recommendations.
To address this, we implemented instant merging of overlapping differences, using developer-edited code blocks as units to create cohesive recommendations aligned with real-world editing behaviors.
\begin{table}[htbp]
    \centering
    \small
    \caption{An example comparison between the diff formats of \texttt{difflib} and \mymodel highlights that \mymodel's format offers higher information density with precise line numbers. }
    \label{tab:simple}
    \begin{tabular}{l|l}
\hline
\multicolumn{1}{c|}{\texttt{difflib} format} & \multicolumn{1}{c}{\mymodel diff format} \\ \hline
\multirow{6}{*}{\begin{tabular}[c]{@{}l@{}}@@ -1,3 +1,3 @@\\ -def Hello()\\ +def GoodBye()\\    print("Say")\\ -  print("Hello")\\ +  print("GoodBye")\end{tabular}} & \multirow{6}{*}{\begin{tabular}[c]{@{}l@{}}1-| def Hello()\\ 1+| def GoodBye()\\ 2 |   print("Say")\\ 3-|   print("Hello")\\ 3+|   print("GoodBye")\end{tabular}} \\&  \\&    \\&   \\& \\&
 \\ \hline
\end{tabular}
\end{table}

\begin{figure}[t]
    \centering
    \includegraphics[width=0.9\linewidth]
    {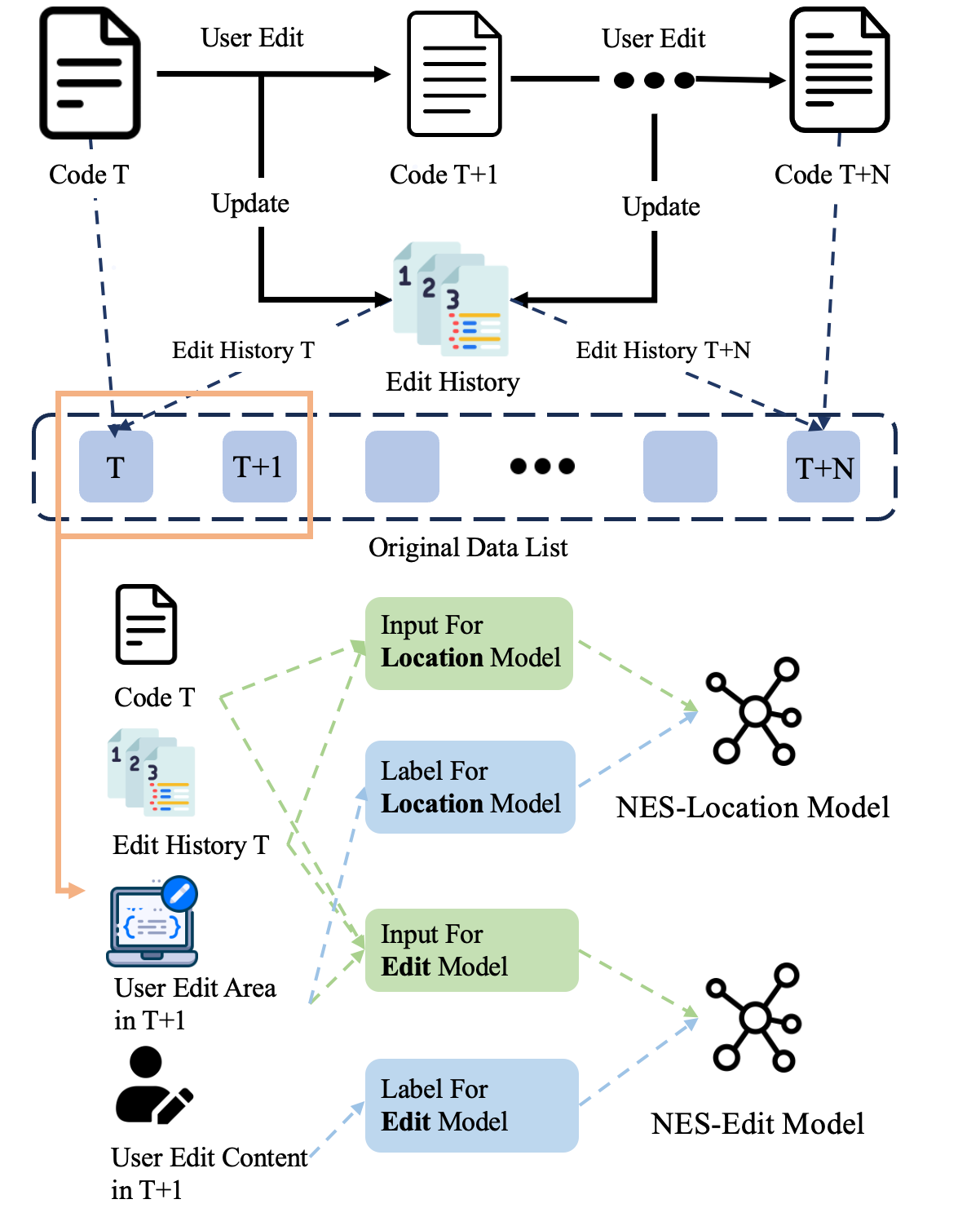}
    \caption{Training dataset collection process}
    \label{fig:training}
    \vspace{-2ex}
\end{figure}

\subsection{Training Dataset Construction\label{sec:datacons}}

Building a high-quality training dataset is key to NES's success. Using developer editing trajectories (Section~\ref{Incremental Difference Detection}), we convert raw data into structured instances for our dual-model architecture through two stages: instance formulation and relevance-based filtering.

\subsubsection{Formulation of Sequential Edit Instances.}
Our process begins by segmenting the continuous stream of code modifications into discrete, sequential training instances. As illustrated in Figure~\ref{fig:training}, we consider any two adjacent timestamps, $T$ and $T+1$, where a developer's action (e.g., an edit) creates a state transition. Each transition yields a data tuple containing:
\begin{itemize}[leftmargin=*]
    \item The pre-edit code state ($C_T$).
    \item The historical editing trajectory up to that point ($H_T$).
    \item The ground-truth edit, comprising the location ($L_{\text{gt}}$) and content ($E_{\text{gt}}$) of the modification performed between $T$ and $T+1$.
\end{itemize}

From this tuple, we formulate training samples:
\begin{itemize}[leftmargin=*]
    \item \textbf{For the NES-Location Model:} The input is the pair ($C_T$, $H_T$), and the target label is the subsequent edit location, $L_{\text{gt}}$.
    \item \textbf{For the NES-Edit Model:} The input consists of ($C_T$, $H_T$) and the ground-truth edit location $L_{\text{gt}}$. The target label is the new code content, $E_{\text{gt}}$.
\end{itemize}

In this manner, we established our preliminary training dataset.

\subsubsection{Relevance-Based Filtering and Negative Sampling.}
A significant challenge with raw trajectory data is the prevalence of spurious correlations. Developers' edit sequences are not always causally linked; a developer might complete one task and immediately start an unrelated one. Training on such uncorrelated sequences could lead the model to learn erroneous associations, resulting in disruptive and irrelevant suggestions.

To mitigate this, we introduce a crucial data refinement stage. We employ an LLM, like Qwen, as a relevance filter to analyze the semantic and causal relationship between the historical edits ($H_T$) and the current edit ($L_{\text{gt}}$, $E_{\text{gt}}$). Based on the LLM's assessment, each instance is processed as follows:
\begin{itemize}[leftmargin=*]
    \item \textbf{Relevant Edits:} If the LLM filter determines that the current edit is a logical and predictable continuation of the historical trajectory, the instance is treated as a \textbf{modification (\texttt{-do})} sample. The original ground-truth edit is retained, teaching the model that the history provides a valid basis to predict the corresponding modification or navigation.
    \item \textbf{Irrelevant Edits:} Conversely, if the filter assesses the edit as uncorrelated with the history, it implies that the developer's next action cannot be reliably predicted from the preceding edits. In this case, the instance is repurposed as a \textbf{preservation (\texttt{-keep})} sample. This crucial step teaches the model to refrain from making a suggestion when the developer’s intent is not inferable from the available history.
\end{itemize}

This filtering and relabeling process transforms the preliminary training dataset into a final, optimized training set. This refined dataset not only enhances the quality of modification-oriented samples but, critically, also incorporates explicit negative samples. This dual focus is essential for training a model that is both proactive in its assistance and discerning enough to avoid generating noise.

Finally, it is essential to elucidate the language distribution of our dataset. Due to the characteristics of our data source developers, the training data exclusively consists of two frontend languages: TypeScript and TypeScriptReact. Nevertheless, as demonstrated in Section~\ref{sec:exp}, our model demonstrates a certain degree of generalization capability on backend languages such as Java and Python. 
\subsection{Model Training}

Our training strategy for the NES-Location and NES-Edit models follows a two-stage process to optimize performance and align with developer preferences. We adapt three pre-trained, small-sized code LLMs for our tasks: Qwen3-4B, Qwen2.5-Coder-7B, and Seed-Coder-8B-Instruct. The training process involves: \textbf{(1) SFT} to establish foundational task capabilities, and \textbf{(2) Reinforcement Learning with DAPO} to further refine the model's behavior based on task-specific rewards.

\subsubsection{Stage 1: SFT}

\begin{figure}[t]
    \centering
    \includegraphics[width=1\linewidth]{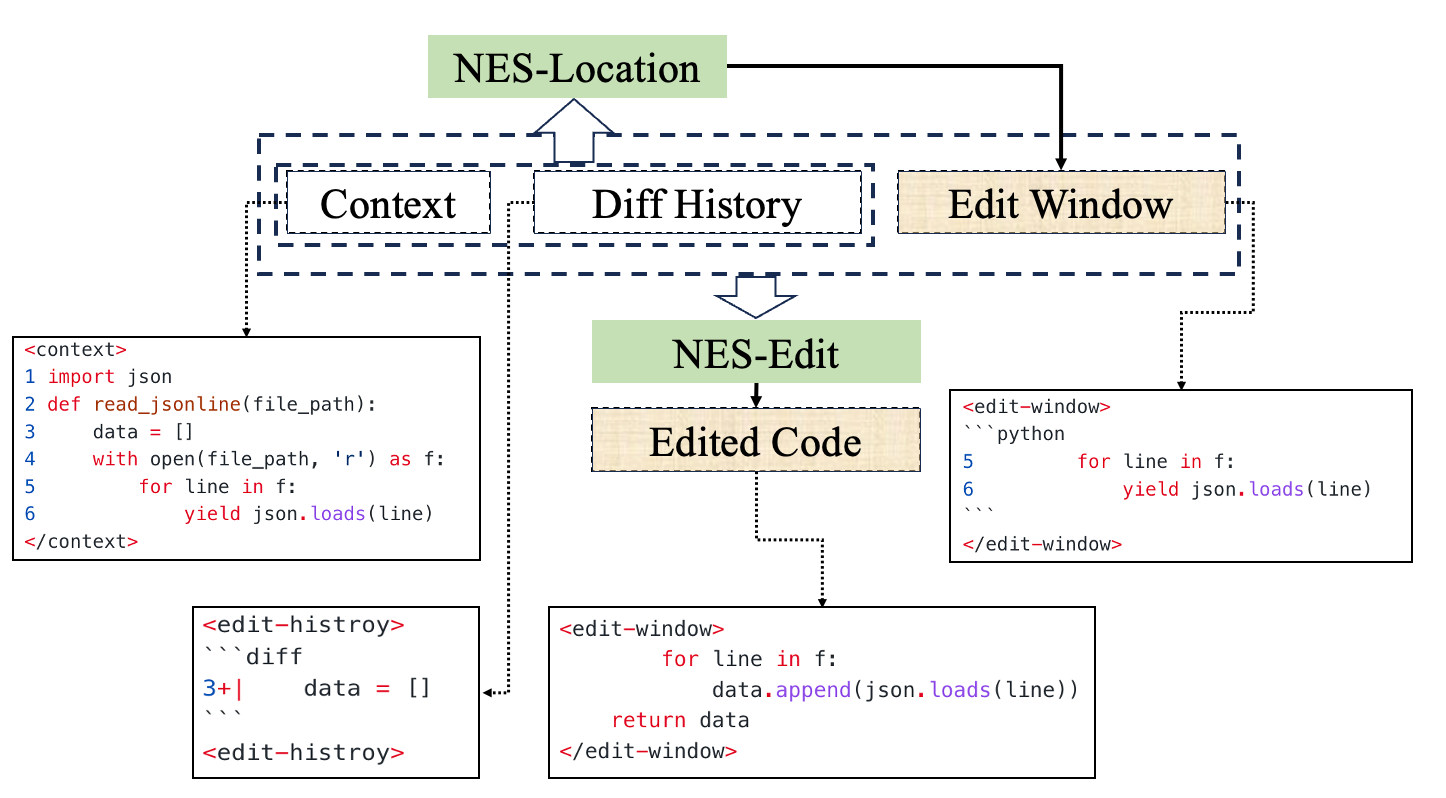}
    \caption{Input-Output Format}
    \label{fig:in_and_out}
\end{figure}

The initial stage uses \emph{SFT} to adapt the base code LLMs to our specific tasks. The goal of this stage is to teach the models the fundamental input-output formats required for next edit suggestion (as shown in Figure~\ref{fig:in_and_out}). We use our large-scale historical editing datasets (\textasciitilde200k samples for \mymodel-Edit, \textasciitilde60k for \mymodel-Location) for this purpose. During SFT, the model learns to process a structured input, which includes the $C_T$, $H_T$, and the $L_{\text{gt}}$ (for the \mymodel-Edit model), to generate a plausible output---either $L_{T+1}$ or $E_{T+1}$. This step establishes a robust baseline model with a solid understanding of code context and common editing patterns.

\subsubsection{Stage 2: Reinforcement Learning with DAPO}


SFT teaches patterns, but lacks task-specific optimization. RL with \emph{DAPO} refines the model by rewarding correct and accurate outputs.
In the RL phase, a custom \emph{reward function} evaluates generated answers ($L_{\text{gen}}$ or $E_{\text{gen}}$) against ground truth based on format correctness and content accuracy, guiding the model toward well-formed and precise outputs.

\paragraph{Reward Function for NES-Location Model}
In contrast, the task of the NES-Location model is to predict a discrete, precise location. In this context, the concept of partial credit is less applicable. Therefore, the reward function $R_{\text{Location}}$ is a straightforward binary signal based purely on accuracy:
\begin{equation}
R_{\text{Location}}(L_{\text{gen}}, L_{\text{gt}}) =
\begin{cases}
    1.0 & \text{if } L_{\text{gen}} \text{ = } L_{\text{gt}} \\
    -1.0 & \text{otherwise}
\end{cases}
\end{equation}

\emph{Reward Function for NES-Edit Model. }
The reward function for the NES-Edit model is designed to prioritize perfect accuracy while also assigning partial credit to semantically similar, potentially useful suggestions. This is achieved through a hierarchical reward structure based on Exact Match Rate (EMR) and Edit Similarity (ES). The reward $R_{\text{Edit}}$ is defined as:
\begin{equation}
R_{\text{Edit}}(E_{\text{gen}}, E_{\text{gt}}) =
\begin{cases}
    1.0 & \text{if } E_{\text{gen}} \text{ = } E_{\text{gt}} \\
    0.5 \times \text{ES}(E_{\text{gen}}, E_{\text{gt}}) & \text{if } \text{ES}(E_{\text{gen}}, E_{\text{gt}}) > 0.5 \\
    -1.0 & \text{otherwise}
\end{cases}
\end{equation}
A reward of \textbf{1.0} is granted for an exact match, strongly encouraging the model to pursue perfect accuracy. If the generated edit is not an exact match but is substantially similar to the ground truth (ES > 0.5), a proportional reward is assigned. The scaling factor of 0.5 ensures this partial credit is strictly less than the reward for a perfect solution, while still guiding the model to generate semantically relevant code. A penalty of \textbf{-1.0} is given for suggestions that are neither correct nor sufficiently similar, preventing the model from learning to produce irrelevant outputs.

This clear and unambiguous reward signal effectively trains the model to maximize its accuracy in predicting the developer's next point of focus. By applying DAPO with these carefully designed reward functions, we enhance the models' practical utility beyond what can be achieved with SFT alone.

\subsection{Optimization of Model Inference\label{optimization}}
The \mymodel framework demands low latency to ensure a seamless developer experience. To tackle this, we implemented various inference optimization techniques during deployment:

\begin{itemize}[leftmargin=*]
\item \emph{Efficient Model Deployment Based on vLLM:}
Used the vLLM framework, leveraging Paged Attention and Dynamic Batching to optimize LLM inference for performance, efficiency, and usability.
\item \emph{Speculative Decoding:}
Applied Speculative Decoding~\cite{Speculative} to pre-generate candidate code segments with a lightweight n-gram model, enabling the \mymodel-Edit Model to verify or refine them, reducing computational overhead.
\item \emph{Prefix Caching:}
Reused intermediate results by caching Key-Value (KV) Cache from historical requests, improving inference efficiency for repetitive prefixes.
\end{itemize}

\section{Evaluation\label{sec:exp}}
We evaluate \mymodel by addressing the following questions:
\begin{itemize}[leftmargin=*]
\item \textbf{RQ1 Next Location Prediction:} To what extent can \mymodel accurately predict the position of the next edit?

\item \textbf{RQ2 Code Edit Generation:} How effectively can \mymodel generate precise edit suggestions?

\item \textbf{RQ3 Model Performance:} Does \mymodel demonstrate sufficient computational efficiency to provide low-latency suggestions in interactive development environments?




\end{itemize}
\subsection{Experiment Setup}

\subsubsection{Benchmarks}

To evaluate the performance of \mymodel, we construct a benchmark of 4 programming languages (i.e., Java, Python, TypeScript and TypeScriptReact) from 1000 open-source and industrial projects. For the open-source projects, we selected the top 100 repositories on GitHub based on star count, while the industrial projects were chosen based on their update frequency. Following the same methodology used for training data, we constructed a unified test dataset comprising 1,000 samples per language. This dataset is systematically divided into two distinct scenarios: \emph{Modification} and \emph{Preservation}.

\begin{itemize}[leftmargin=*]
    \item \textbf{} Modification (\texttt{-do}): 500 samples per language where subsequent edits occur in a different location, testing the model's predictive navigation capability.
    \item \textbf{} Preservation (\texttt{-keep}): 500 samples per language where edits remain in the same location or terminate, assessing the model's ability to avoid unnecessary navigation suggestions.
\end{itemize}

\subsubsection{Metrics}

For RQ1, which evaluates the prediction of the next editing position, the primary metric used is \emph{Accuracy (ACC)}, measuring whether the predicted edit location matches the ground truth.

For RQ2, which evaluates code-editing performance, the three key metrics are: \emph{Edit Similarity (ES)}~\cite{Song_2024} for semantic alignment between the generated and target edits; \emph{Exact Match Rate (EMR)} for syntactic precision of the generated edits for the modification task; and \emph{Accuracy (ACC)} for correctly predict preservation.

For RQ3, which investigates model inference speed, the performance metrics are input and output \emph{tokens/s} speed.

\subsubsection{Baseline Models}
We compare \mymodel to the models.

\begin{itemize}[leftmargin=*]
    \item Qwen~\cite{qwen3technicalreport}, developed by Alibaba Cloud, is a renowned LLM, while Qwen-Coder~\cite{Yang2024Qwen25TR} specializes in programming and software development tasks.
    \item Claude~\cite{TheC3}, created by Anthropic, has strong capabilities in coding and can assist with a wide range of programming tasks, including writing, debugging, and explaining code.
    \item Seed-Coder~\cite{Seed2025SeedCoderLT}, recently released by ByteDance, is a lightweight yet powerful open-source code-focused large language model (LLM) that delivers strong performance across a variety of coding benchmarks.
    \item Zeta~\cite{zeta}, recently introduced by Zed, is an advanced tool designed to provide instruction-free edit generation capabilities. Zeta intuitively generates context-aware code or text modifications based solely on the input provided.
    \item CoEdPilot~\cite{Liu2024CoEdPilotRC} jointly trains a localization LLM (to pinpoint edits) and a generation LLM. However, the reliance on explicit human instructions constrains their overall editing efficiency.
\end{itemize}

The models with relatively smaller sizes include Qwen3-4B, Qwen2.5-Coder-7B, and Seed-Coder-8B-Instruct. These models will undergo supervised fine-tuning (SFT) using the \mymodel SFT and DAPO datasets. Zeta-7B, however, was not subjected to additional training, as it is already built on Qwen2.5-Coder-7B and incorporates Zeta's own training data. Instead, we use Zeta-7B as a baseline.

The models with relatively larger sizes including Qwen3-32B, Qwen2.5-Coder-32B-Instruct, Claude3, and Claude4. These were not trained further due to the following reasons:
a. Large models are challenging to meet the low-latency requirements in code editing scenarios.
b. Training large models is resource-intensive and consumes significant computational resources.

We select CoEdPilot~\cite{Liu2024CoEdPilotRC} as the baseline from the family of explicit human-instruction-based code editing tools, alongside models like CodeT5~\cite{Wang2021CodeT5IU}, CoditT5~\cite{Zhang2022CoditT5PF}, Coeditor~\cite{wei2024coeditor}, EfficientEdit~\cite{wang2025reusegenerateacceleratingcode} and so on.
As our evaluation environment does not natively support CoEdPilot's required human instruction input format, we left the corresponding input field blank during testing.




\subsubsection{Settings} In our testing environment, we utilized a single NVIDIA A100-80GB GPU for evaluation purposes. For the production deployment in our online real-world environment, we implemented a configuration of NVIDIA L20-48GB GPUs. Both setups were based on vLLM version 0.8.5.

\subsection{Results}

\begin{table*}[htbp]
\fontsize{9}{11}\selectfont 
    \centering
    \caption{Accuracy comparison for location tasks (-do and -keep) across models. ``model+SFT'' and ``model+SFT+DAPO'' indicate training on the base model with the \mymodel SFT dataset alone and on both the SFT and DAPO datasets.}
    \label{location:location_comparison_combined}
    \begin{tabular}{l|cccccccc|cc}
        \hline
        \multicolumn{1}{c|}{\multirow{2}{*}{Model}} & \multicolumn{2}{c}{TypeScript} & \multicolumn{2}{c}{TypeScriptReact} & \multicolumn{2}{c}{Java} & \multicolumn{2}{c|}{Python} & \multicolumn{2}{c}{Average} \\
        \cline{2-11} 
        \multicolumn{1}{c|}{} & do & keep & do & keep & do & keep & do & keep & do & keep \\
        \hline
        Qwen2.5-Coder-7B  & 8.7\% & 2.2\% & 9.8\% & 6.8\% & 10.3\% & 1.9\% & 11.8\% & 3.7\% & 10.1\% & 3.7\% \\ \hline
        Seed-Coder-8B-Instruct & 8.0\% & 1.9\% & 9.0\% & 3.7\% & 6.5\% & 1.8\% & 13.5\% & 4.1\% & 9.3\% & 2.9\% \\ \hline
        Qwen3-4B & 5.6\% & 1.8\% & 4.4\% & 2.0\% & 7.7\% & 0.8\% & 8.7\% & 3.0\% & 6.6\% & 1.9\% \\ \hline
        Qwen3-32B & 51.6\% & 6.7\% & 46.4\% & 3.1\% & 64.2\% & 1.0\% & 63.0\% & 2.9\% & 56.3\% & 3.4\% \\ \hline
        Qwen2.5-Coder-32B-Instruct & 15.6\% & 15.5\% & 12.8\% & 17.1\% & 12.9\% & 15.7\% & 27.2\% & 18.2\% & 17.1\% & 16.6\% \\ \hline
        Claude3 & 57.2\% & 5.4\% & 54.2\% & 3.8\% & 63.6\% & 21.0\% & 66.4\% & 3.4\% & 60.4\% & 8.4\% \\ \hline
        Claude4 & 65.2\% & 8.0\% & 65.0\% & 10.0\% & 60.0\% & 15.6\% & 77.6\% & 4.7\% & 67.0\% & 9.6\% \\ \hline
        CoEdPilot & 63.2\% & 4.4\% & 62.6\% & 3.0\% & 12.8\% & 6.4\% & 17.6\% & 27.4\% & 31.6\% & 8.2\% \\
        \hline \hline
        Qwen2.5-Coder-7B+SFT & 62.4\% & \textbf{94.6\%} & 69.3\% & \textbf{97.0\%} & 51.3\% & \textbf{95.3\%} & 68.4\% & 89.3\% & 62.8\% & 94.0\% \\ \hline
        Seed-Coder-8B-Instruct+SFT & 73.2\% & \textbf{94.6\%} & 75.7\% & 96.6\% & 64.1\% & 90.4\% & 77.4\% & \textbf{94.8\%} & 72.6\% & \textbf{94.1\%} \\ \hline
        Qwen3-4B+SFT & 74.6\% & 90.5\% & 76.9\% & 91.6\% & 64.1\% & 89.3\% & 74.6\% & 86.0\% & 72.6\% & 89.4\% \\ \hline \hline
        Qwen2.5-Coder-7B+SFT+DAPO & 66.5\% & 85.3\% & 71.7\% & 89.4\% & 53.6\% & 86.3\% & 70.4\% & 85.1\% & 65.6\% & 86.5\% \\ \hline
        Seed-Coder-8B-Instruct+SFT+DAPO & \textbf{77.6\%} & 85.1\% & \textbf{78.9\%} & 87.8\% & \textbf{73.9\%} & 83.6\% & \textbf{80.4\%} & 83.6\% & \textbf{77.7\%} & 85.0\% \\ \hline
        Qwen3-4B+SFT+DAPO & 76.5\% & 82.3\% & 77.3\% & 86.2\% & 71.8\% & 80.1\% & 76.9\% & 77.8\% & 75.6\% & 81.6\% \\ \hline
    \end{tabular}
    \par\noindent\parbox{\linewidth}{{\footnotesize * Training large models on more data was infeasible because of the necessary \textbf{low-latency performance} and \textbf{restrictive training resource constraints}.
    \textbf{Zeta} was not evaluated because it lacks navigation functionality. \textbf{CoEdPilot} was evaluated without human instructions.} }
\end{table*}


\begin{table*}[htbp]
\fontsize{9}{11}\selectfont 
    \centering
    \caption{Comparison of location prediction performance with varying max edit history lengths on the Qwen3-4B+SFT model}
    \label{location:diff_num}
    
    \begin{tabular}{c|cccccccc|cc}
        \hline
        \multirow{2}{*}{\begin{tabular}[c]{@{}c@{}}Max Edit\\ History\end{tabular}} & \multicolumn{2}{c}{TypeScript} & \multicolumn{2}{c}{TypeScriptReact} & \multicolumn{2}{c}{Java} & \multicolumn{2}{c|}{Python} & \multicolumn{2}{c}{Average} \\
        \cline{2-11}
         & do & keep & do  & keep  & do  & keep  & do  & keep & do & keep \\
        \hline
        1 & 73.7\% & 84.1\% & 73.9\% & 89.4\% & \textbf{69.2\%} & 84.7\% & \textbf{75.9\%} & 84.9\% & \textbf{73.2\%} & 85.8\% \\ \hline
        3 & \textbf{74.6\%} & \textbf{90.5\%} & \textbf{76.9\%} & \textbf{91.6\%} & 64.1\% & \textbf{89.3\%} & 74.6\% & \textbf{86.0\%} & 72.6\% & \textbf{89.4\%} \\ \hline
        5 & 68.7\% & 80.4\% & 70.5\% & 82.6\% & 56.4\% & 78.8\% & 73.2\% & 76.1\% & 67.2\% & 79.5\% \\ \hline
        7 & 68.3\% & 79.8\% & 70.6\% & 81.8\% & 56.4\% & 78.8\% & 74.1\% & 76.1\% & 67.4\% & 79.1\% \\ \hline
        9 & 67.6\% & 82.2\% & 71.4\% & 86.2\% & 60.0\% & 80.1\% & 72.1\% & 75.7\% & 67.8\% & 81.0\% \\ \hline
    \end{tabular}
\end{table*}

\subsubsection{RQ1 \mymodel-Location:} \emph{To what extent can \mymodel~accurately predict the location of the next edit?}

Table~\ref{location:location_comparison_combined} demonstrates the accuracy of the \mymodel-Location model in two distinct scenarios: \textbf{modification (\texttt{-do})} and \textbf{preservation (\texttt{-keep})}. The terms ``model+SFT'' and ``model+SFT+DAPO'' indicate training on the base model with the \mymodel SFT dataset alone and on both the \mymodel SFT and DAPO datasets, respectively.

Before post-training, the best base model, Claude4, achieves 67.0\% and 9.6\% average accuracy on the \texttt{-do} and \texttt{-keep} tasks, respectively. Models with small size, such as Qwen2.5-Coder-7B, perform poorly, with only 10.1\% and 3.7\% on the \texttt{-do} and \texttt{-keep} tasks, respectively.
\mymodel SFT dataset significantly improves models with small size. For \texttt{-do}, Qwen2.5-Coder-7B jumps from 10.1\% to 62.8\%, Seed-Coder-8B-Instruct from 9.3\% to 72.6\%, and Qwen3-4B from 6.6\% to 72.6\%. For \texttt{-keep}, Qwen2.5-Coder-7B rises from 3.7\% to 94.0\%, Seed-Coder-8B-Instruct from 2.9\% to 94.1\%, and Qwen3-4B from 1.9\% to 89.4\%. 

\emph{Limits.} Due to constraints on \textbf{low-latency requirements} and \textbf{training resources}, large models were not trained on the SFT or DAPO dataset. Furthermore, since \textbf{Zeta} lacks navigation functionality, its results are excluded from Table~\ref{location:location_comparison_combined}. Finally, \textbf{CoEdPilot}'s reliance on explicit human inputs explains why its performance did not surpass that of other tools. A similar result was observed in the subsequent evaluation.

\noindent\begin{tcolorbox}[size=title, opacityfill=0.1, nobeforeafter]
\textbf{Impact of SFT:} The \mymodel SFT dataset significantly boosts model performance in predicting the next editing position, improving Qwen3-4B's accuracy from 6.6\% to 72.6\% on the \texttt{-do} task and from 1.9\% to 89.4\% on the \texttt{-keep} task.
\end{tcolorbox}


Training smaller models on the DAPO dataset results in a slight but notable improvement in average navigation prediction (\texttt{-do}) accuracy: Qwen2.5-Coder-7B improves from 62.8\% to 65.6\%, Seed-Coder-8B-Instruct increases from 72.6\% to 77.7\%, and Qwen3-4B rises from 72.6\% to 75.6\%. But, on the \texttt{-keep} task, Qwen2.5-Coder-7B declines from 94.0\% to 86.5\%, Seed-Coder-8B-Instruct declines from 94.1\% to 85.0\%, and Qwen3-4B drops from 89.4\% to 81.6\%.

The DAPO-trained model tends to suggest location changes more frequently. This is because changing is more frequent in the real-world development practices. 
We accept the minor drop in \texttt{-keep} performance in favor of improved \texttt{-do} accuracy. DAPO was chosen because predicting necessary jumps accurately matters more than avoiding suggestions in developers' workflow.


\noindent\begin{tcolorbox}[size=title, opacityfill=0.1, nobeforeafter]
\textbf{Impact of DAPO:} 
The DAPO-trained model tends to suggest location changes more frequently, leading to a slight drop in \texttt{-keep} task accuracy. However, 
the \mymodel-Location model was chosen for its superior \texttt{-do} task performance, as location changes are more prevalent in practice.
\end{tcolorbox}

Table~\ref{location:diff_num} demonstrates the impact of historical context length (\texttt{Max Edit History}) on Qwen3-4B+SFT. A history length of 3 achieves the best balance, with 72.6\% \texttt{-do} average accuracy and 89.4\% \texttt{-keep} average accuracy. Short histories (e.g., 1) lack sufficient context for complex intents, while longer histories (e.g., 5+) introduce irrelevant noise, hindering accuracy. 

\noindent\begin{tcolorbox}[size=title, opacityfill=0.1, nobeforeafter]
\textbf{Impact of Edit History Length:} Short histories may lack context, while long histories can introduce noise, both hindering accuracy. The results show that a history length of 3 is optimal.
\end{tcolorbox}

{    
\begin{table*}[htbp]
    \centering
\fontsize{9}{11}\selectfont
    \caption{Similarity comparison for modification tasks (-do). ``model+SFT'' and ``model+SFT+DAPO'' indicate training on the base model with the \mymodel SFT dataset alone and on both the SFT and DAPO datasets.}
    \label{tab:model_comparison_do}

    \begin{tabular}{l|cccc|c}
\hline  \multirow{2}{*}{Model}    & \multicolumn{1}{c}{TypeScript}                                  & \multicolumn{1}{c}{TypeScriptReact}                             & \multicolumn{1}{c}{Java}              & \multicolumn{1}{c|}{Python}                                 & Average                                                          \\ \cline{2-6}
\multicolumn{1}{l|}{}                           & \multicolumn{1}{c}{ES / EMR}                                    & \multicolumn{1}{c}{ES / EMR}                                    & \multicolumn{1}{c}{ES / EMR}          & \multicolumn{1}{c|}{ES / EMR}                               & ES / EMR                                                         \\ \hline
\multicolumn{1}{l|}{Qwen2.5-Coder-7B}  & 72.17/12.4\%  & 69.52/6.4\%  & 64.22/4.6\% & \multicolumn{1}{c|}{72.52/8.8\%} & 69.61/8.1\%  \\ \hline
\multicolumn{1}{l|}{Seed-Coder-8B-Instruct} & 71.30/14.9\%  & 69.35/7.0\%  & 69.91/8.8\% & \multicolumn{1}{c|}{70.06/9.8\%}  & 70.16/10.1\%   \\ \hline
\multicolumn{1}{l|}{Qwen3-4B} & 76.77/6.4\% & 80.46/13.5\%  & 73.32/6.6\%  & \multicolumn{1}{c|}{77.23/7.0\%} & 76.94/8.5\% \\ \hline
\multicolumn{1}{l|}{Qwen3-32B} & 79.32/16.7\% & 76.81/11.4\% & 75.66/13.8\% & 79.01/15.2\% & 77.70/14.3\%  \\ \hline
\multicolumn{1}{l|}{Qwen2.5-Coder-32B-Instruct}& 81.33/19.6\%& 81.34/13.2\%  & 76.39/14.0\%  & 79.80/19.0\%  & 79.72/16.5\%  \\ \hline
\multicolumn{1}{l|}{Zeta} & 86.22/17.3\% & 86.29/9.2\% & 86.34/15.0\% & \multicolumn{1}{c|}{84.90/8.8\%} & 85.94/12.6\% \\ \hline
\multicolumn{1}{l|}{Claude3}& 84.81/24.3\% & 89.0/19.2\% & 82.81/18.8\% & 81.72/22.8\% &  84.59/21.3\% \\ \hline
\multicolumn{1}{l|}{Claude4}  & 85.37/25.7\%  & 85.56/19.6\% & 83.40/22.4\% & \multicolumn{1}{c|}{83.38/24.8\%} & 84.43/23.1\%    \\ \hline 
CoEdPilot & 72.34/4.3\% & 77.73/1.6\% & 72.06/0.8\% & 60.88/1.8\% & 70.75/2.1\%  \\ \hline \hline
\multicolumn{1}{l|}{Qwen2.5-Coder-7B+SFT} & 92.49/28.8\% & 92.63/28.8\% & 89.17/19.6\%       & \multicolumn{1}{c|}{89.61/22.6\%}  & 90.98/25.0\% \\ \hline
\multicolumn{1}{l|}{Seed-Coder-8B-Instruct+SFT} & 92.29/30.2\% & 92.85/34.0\% & 89.55/20.4\%       & \multicolumn{1}{c|}{89.54/22.0\%}  & 91.06/26.7\% \\ \hline
\multicolumn{1}{l|}{Qwen3-4B+SFT} & 92.03/27.0\% & 90.69/29.6\% & 89.29/20.2\% & \multicolumn{1}{c|}{89.30/21.6\%} & 90.33/24.6\% \\ \hline \hline
\multicolumn{1}{l|}{Qwen2.5-Coder-7B+SFT+DAPO} & 91.82/\textbf{32.4\%} & 92.67/\textbf{34.4\%} & 89.99/21.0\%       & \multicolumn{1}{c|}{90.03/25.4\%}  & 91.13/28.3\% \\ \hline
\multicolumn{1}{l|}{Seed-Coder-8B-Instruct+SFT+DAPO} & 91.90/31.4\% & 93.21/34.2\% & \textbf{90.15}/\textbf{23.6}\%       & \multicolumn{1}{c|}{\textbf{90.08}/\textbf{26.0\%}}  & 91.34/\textbf{28.8\%} \\ \hline
\multicolumn{1}{l|}{Qwen3-4B+SFT+DAPO} & \textbf{92.77}/30.6\% & \textbf{93.28}/34.2\% & 89.97/21.4\% & \multicolumn{1}{c|}{89.41/24.6\%} & \textbf{91.36}/27.7\% \\ \hline
\end{tabular}
\end{table*}

\begin{table*}[htbp]
\fontsize{9}{11}\selectfont 
    \centering
    \caption{Accuracy comparison for preservation tasks (-keep). ``model+SFT'' and ``model+SFT+DAPO'' indicate training on the base model with the \mymodel SFT dataset alone and on both the SFT and DAPO datasets.}
    \label{tab:model_comparison_keep}
    \begin{tabular}{l|cccc|c}
\hline
        \multirow{1}{*}{Model}   & \multicolumn{1}{c}{TypeScript} & \multicolumn{1}{c}{TypeScriptReact} & \multicolumn{1}{c}{Java} & Python          & Average         \\ \hline
Qwen2.5-Coder-7B & 11.6\% & 8.0\% & 4.6\%  & 8.8\%  & 8.3\%          \\ \hline
Seed-Coder-8B-Instruct  & 2.6\%   & 3.2\%     & 2.0\%      & 3.8\%  & 2.9\%          \\ \hline
Qwen3-4B & 35.2\% & 37.4\% & 28.8\%  & 44.8\%  & 36.6\%          \\ \hline
Qwen3-32B & 14.0\%   & 21.2\%   & 9.4\%  & 18.8\%  & 15.9\%   \\ \hline
Qwen2.5-Coder-32B-Instruct & 11.0\%   & 11.0\%   & 7.2\%  & 24.2\%  & 13.4\%   \\ \hline
Zeta  & 40.0\%   & 35.4\%     & 43.2\%      & 44.2\%  & 40.7\%          \\ \hline
Claude3 & 21.0\%   & 17.6\%   & 24.0\%  & 19.4\%  & 20.5\%   \\ \hline
Claude4                   & 15.6\%                          & 15.2\%                               & 11.0\%                      & 17.0\%            & 14.7\%  \\    \hline   
CoEdPilot & 11.2\% & 4.6\% & 7.0\% & 23.2\% & 11.5\%  \\ \hline \hline 
Qwen2.5-Coder-7B+SFT & 94.4\%   & 90.2\%   & 87.4\%  & 92.0\%  & 91.0\%   \\ \hline
Seed-Coder-8B-Instruct+SFT & 92.2\%   & 86.0\%   & 88.6\%  & 89.8\%  & 89.2\%          \\ \hline
\multicolumn{1}{l|}{Qwen3-4B+SFT} & 92.4\% & 84.8\% & 88.0\% & 88.8\% & 88.5\% \\ \hline \hline
Qwen2.5-Coder-7B+SFT+DAPO & 84.8\%   & 85.2\%   & 80.8\%  & 86.2\%  & 84.3\%   \\ \hline
Seed-Coder-8B-Instruct+SFT+DAPO & 85.2\%   & 80.0\%   & 78.2\%  & 83.4\%  & 81.7\%   \\ \hline
\multicolumn{1}{l|}{Qwen3-4B+SFT+DAPO} & 86.2\% & 88.8\% & 82.0\% & 88.2\% & 86.3\% \\ \hline
\end{tabular}
\end{table*}

\begin{table*}[htbp]
\fontsize{9}{11}\selectfont 
    \centering
    \caption{Similarity comparison with different ratios of unmodified data}
    \label{tab:unmodified_data_ratio}
    \begin{tabular}{c|cccccccc|cc}
    \hline
Ratio of   & \multicolumn{2}{c}{TypeScript} & \multicolumn{2}{c}{TypeScriptReact} & \multicolumn{2}{c}{Java} & \multicolumn{2}{c|}{Python} & \multicolumn{2}{c}{Average} \\ \cline{2-11} 
        Unmodified & do & keep & do & keep & do & keep & do & keep     & do & keep       \\ \cline{2-11} 
        Data  & ES / EMR & Acc & ES / EMR & Acc & ES / EMR & Acc & ES / EMR & Acc  & ES / EMR  & Acc        \\ \hline
        10\% & 91.44/26.7\% & 92.2\% & 91.22/21.2\% & 87.0\% & 89.03/18.8\% & 87.6\% &  88.95/21.6\% & 89.6\% & 90.16/22.1\% & 89.1\%\\
        \hline
        20\% & 91.83/26.1\% & 93.4\% & 91.36/20.6\% & 89.6\% & 89.06/16.4\% & 90.2\% & 89.05/20.4\% & 92.2\% & 90.33/20.9\% & 91.4\%\\
        \hline
        30\% & 91.73/24.9\% & 94.8\% & 90.81/20.4\% & 90.8\%& 88.11/15.4\% & 89.8\% & 89.19/19.6\% & 92.6\%  & 89.96/20.1\% & 92.0\%\\
        \hline
        40\% & 91.55/24.1\% & 96.4\% & 91.20/20.4\% & 92.0\% & 89.17/17.2\% & 91.6\% & 89.22/19.6\% & 94.0\% & 90.28/20.33\% & 93.5\%\\
        \hline
        50\% & 91.61/24.3\% & 96.4\% & 90.87/18.6\% & 94.6\% &  88.72/14.2\% & 93.9\% & 89.40/19.2\% & 95.4\% & 90.15/19.1\% & 95.1\%\\
        \hline
        60\% & 91.89/21.4\% & 98.6\% & 91.01/20.0\% & 96.0\% & 88.73/12.2\% & 95.4\% & 88.92/15.6\% & 95.8\% & 90.14/17.3\% & 96.5\%\\
        \hline
    \end{tabular}
\end{table*}

\begin{table*}[htbp]
\fontsize{9}{11}\selectfont 
    \centering
    \caption{Comparison of code edit generation with varying maximum edit history lengths on the Qwen3-4B+SFT model}
    \label{tab:diff_num_edit}
    \begin{tabular}{c|cccccccc|cc}
        \hline
        Max & \multicolumn{2}{c}{TypeScript} & \multicolumn{2}{c}{TypeScriptReact} & \multicolumn{2}{c}{Java} & \multicolumn{2}{c|}{Python} & \multicolumn{2}{c}{Average} \\
        \cline{2-11}
        Edit & do & keep & do & keep & do & keep & do & keep  & do & keep\\
        \cline{2-11}
        History & ES / EMR & Acc & ES / EMR & Acc & ES / EMR & Acc & ES / EMR & Acc  & ES / EMR & Acc\\
        \hline
        1 & 96.81/21.7\% & 93.8\% & 91.35/21\% & 85.4\% & 88.88/17.0\% & 88.4\% & 89.16/19.8\% & 91.2\% & 91.55/19.9\% & 89.7\% \\ \hline
        3 & 92.03/27.0\% & 92.4\% & 90.69/29.6\% & 84.8\% & 89.29/20.2\% & 88.0\% & 89.30/21.6\% & 88.8\% & 90.33/24.6\% & 88.5\% \\ \hline
        5 & 91.83/26.1\% & 93.4\% & 91.36/20.6\% & 89.6\% & 89.06/16.4\% & 90.2\% & 89.05/20.4\% & 92.2\% & 90.33/20.9\% & 91.4\% \\ \hline
        7 & 92.49/29.4\% & 95.0\% & 91.43/23.0\% & 90.8\% & 88.13/17.6\% & 89.2\% & 88.06/21.2\% & 91.6\% & 90.03/22.8\% & 91.7\% \\ \hline
        9 & 91.61/29.2\% & 93.0\% & 91.07/23.0\% & 87.8\% & 89.06/16.4\% & 90.2\% & 89.05/20.4\% & 92.2\% & 90.20/22.3\% & 90.8\% \\ \hline
    \end{tabular}
\end{table*}





\subsubsection{RQ2 Code Edit Generation:} \emph{How effectively can \mymodel generate precise edit suggestions?}

Table~\ref{tab:model_comparison_do} and Table~\ref{tab:model_comparison_keep} present the performance of the \mymodel-Edit Model across four programming languages: Java, Python, TypeScript, and TypeScript React.
In the \texttt{-do} task, Zeta achieves the highest ES score of 85.94\%, while Claude4 records the best EMR score of 23.1\%. Following SFT fine-tuning, smaller models demonstrate improved ES and EMR scores. For instance, Qwen2.5-Coder-7B+SFT achieves an ES of 90.98\% and EMR of 25.0\%, Seed-Coder-8B-Instruct+SFT achieves an ES of 91.06\% and EMR of 26.7\%, and Qwen3-4B+SFT achieves an ES of 90.33\% and EMR of 24.6\%.
In the \texttt{-keep} task, Zeta achieves the highest initial accuracy of 40.7\%. After SFT fine-tuning, smaller models show significant improvements in accuracy, with Qwen2.5-Coder-7B+SFT reaching 91.0\%, Seed-Coder-8B-Instruct+SFT achieving 89.2\%, and Qwen3-4B+SFT attaining 88.5\%.

\noindent\begin{tcolorbox}[size=title, opacityfill=0.1, nobeforeafter]
\textbf{Impact of SFT:} The NES SFT dataset greatly improves model performance in code edit generation.
\end{tcolorbox}

Smaller models trained on the DAPO dataset achieve moderate improvements in average code edit generation (\texttt{-do}) similarity scores. For example, Qwen2.5-Coder-7B improves from an ES of 90.98\% to 91.13\% and an EMR of 25.0\% to 28.3\%; Seed-Coder-8B-Instruct increases from an ES of 91.06\% to 91.34\% and an EMR of 26.7\% to 28.8\%; and Qwen3-4B improves from an ES of 90.33\% to 91.36\% and an EMR of 24.6\% to 27.7\%.

For the \texttt{-keep} task, Qwen2.5-Coder-7B experiences a decline in accuracy from 91.0\% to 84.3\%, Seed-Coder-8B-Instruct drops from 89.2\% to 81.7\%, and Qwen3-4B drops from 88.5\% to 86.3\%. The slight \texttt{-keep} drop is due to the DAPO-trained model suggesting changes more frequently, but we deploy it for its alignment with high-frequency practical use cases.


\noindent\begin{tcolorbox}[size=title, opacityfill=0.1, nobeforeafter]
\textbf{Impact of DAPO:} The DAPO-trained model demonstrates improved similarity scores for the \texttt{-do} task, yet shows a minor drop in accuracy for the \texttt{-keep} task. However, we have chosen to adopt the model trained with DAPO, as it better aligns with the high-frequency practices observed in real-world development.
\end{tcolorbox}

We investigated the impact of varying proportions of unmodified data in the training set, analyzing ratios from 10\% to 60\% (Table~\ref{tab:unmodified_data_ratio}). The results show that higher proportions of unmodified data improve performance on the \texttt{-keep}  test sets, while leading to a decline in performance on the \texttt{-do}  test sets. After thorough analysis, a 20\% proportion of unmodified data was determined to be optimal.
As a result, we selected the 20\% unmodified data ratio as the standard configuration for all subsequent experiments.

\noindent\begin{tcolorbox}[size=title, opacityfill=0.1, nobeforeafter]
\textbf{Impact of Unmodified Data:} We selected the 20\% unmodified data ratio as the standard configuration for all subsequent experiments.
\end{tcolorbox}


Table~\ref{tab:diff_num_edit} shows that a history length of 3 is optimal for the \texttt{-do} task (ES 90.33\%, EMR 24.6\%), while a length of 7 yields the best \texttt{-keep} accuracy (91.7\%), suggesting better code edits may require longer historical trajectories.


\noindent\begin{tcolorbox}[size=title, opacityfill=0.1, nobeforeafter]
\textbf{Impact of Edit History Length:} Generating better code edits may require a more extensive historical trajectory.
\end{tcolorbox}

\subsubsection{RQ3 Model Performance:}\emph{Does NES demonstrate sufficient computational efficiency to provide low-latency suggestions in interactive development environments?}




Table~\ref{tab:token_processing_speeds} shows that Prefix Caching (PC)~\cite{zheng2024sglangefficientexecutionstructured}, for reusing prompt computations, and Speculative Decoding (SD)~\cite{Chen2023AcceleratingLL, Speculative}, for faster token generation, significantly boost performance, with SD showing the largest gains. Evaluated on vLLM, \mymodel demonstrates superior input throughput over Zeta, while a latency comparison with CoEdPilot was deemed infeasible, as its non-vLLM-based architecture would require a costly re-implementation for an equitable benchmark. These optimizations enable \mymodel to achieve an average latency of 250ms in live deployment.

{\small
\begin{table}[htbp]
    \centering
    \caption{Comparison of input and output token processing speeds for different model configurations. }
    \label{tab:token_processing_speeds}
    \begin{tabular}{l|c|c}
        \hline
        Model configurations & input tokens/s & output tokens/s \\
        \hline
        \mymodel-Edit+vLLM & 4800 & 71 \\ \hline
        \mymodel-Edit+vLLM+PC & 5750 & 85 \\ \hline
        \mymodel-Edit+vLLM+PC+SD & 8500 & 125 \\ \hline
        Zeta+vLLM+PC+SD & 5400 & 125 \\ \hline
    \end{tabular}
    \par\noindent\parbox{\linewidth}{{\footnotesize * CoEdPilot's non-vLLM architecture is incompatible with PC and SD optimizations.} }
\end{table}
}

\vspace{-3mm}
\noindent\begin{tcolorbox}[size=title, opacityfill=0.1, nobeforeafter]
\textbf{Answer to RQ3:} The SD optimization delivers the most significant improvement in inference input/output throughput, enabling \mymodel to achieve an average inference latency of 250ms.
\end{tcolorbox}

\section{Industrial Deployment}
\label{subsec:industrial_tools}

\subsection{Design Rationale}
In transitioning the \mymodel framework to a scalable, industrial-grade solution, we faced the challenge of balancing model performance, inference latency, and deployment cost. To meet the stringent responsiveness demands of real-time code editing, we chose the \texttt{Qwen3-4B} model as the core engine, post-train it with our SFT and DAPO datasets (see Section~\ref{sec:datacons}) for production deployment. This decision was based on several key considerations:

\begin{itemize}[leftmargin=*]
    \item \emph{Trade-off between Model Size and Latency:} Larger scale models, like 8B-parameter ones, have high computational needs and slower inference, making them unsuitable for real-time code editing's low-latency requirements. The Qwen3-4B model, being smaller, offers faster inference, aligning better with our goal of achieving a "zero-latency" user experience in \mymodel.


    \item \emph{Excellent Cost-Performance Ratio after DAPO Training:} As detailed in Section~\ref{sec:exp}, our two-stage training process greatly improved Qwen3-4B's performance. The \texttt{Qwen3-4B+SFT+DAPO} model ranks among the best small-scale models and rivals larger ones. While DAPO slightly reduces performance on the \texttt{-keep} task, this trade-off is acceptable due to its superior results on the more frequent practical \texttt{-do} task, validating our strategy of leveraging high-quality data and advanced training to balance performance and efficiency in a compact model.




    \item \emph{Feasibility for Large-Scale Deployment:} In industrial applications, ensuring service stability and scalability is critical. We deployed the \texttt{Qwen3-4B+SFT+DAPO} model across a configuration of NVIDIA L20 GPUs with the optimizations in Section~\ref{optimization}, tailored for large-scale deployment. \mymodel consistently achieved an average response time of under 250ms, delivering a seamless user experience while handling tens of thousands of daily calls. These results confirm the model's effectiveness and reliability for industrial use.


\end{itemize}

\subsection{Empirical Evaluation of User Experience}


\mymodel builds upon our prior work, PEACE~\cite{ren2025peaceefficientprojectlevelefficiency}, with repo-level program understanding~\cite{zhang-etal-2025-galla, tao2025codegraphmodelcgm, 10.1145/3691620.3694999}, but its reliance on live user editing data—captured exclusively via our IDE plugin—mandates an in-the-wild evaluation with internal developers.

A direct comparison of acceptance rates is misleading due to fundamental design differences. Tools like CoEdPilot use a "one-to-many" user-description-based model, which naturally produces high raw acceptance rates. In contrast, \mymodel's passive, "one-by-one" suggestion model generates far more frequent interactions. To bridge this gap, we define an Equivalent Acceptance Rate (EAR) by multiplying the raw rate by the average number of lines accepted per event (1.84, based on 6 months of data).

This normalization shows \mymodel's effective rates for \mymodel-Edit and \mymodel-Location tasks are 43.44\% and 51.55\%, respectively. These results confirm that \mymodel offers comparable quantitative performance while providing a superior, description-free user experience. This experiential advantage was further validated in a comparative user study against another tool, Zeta, where \mymodel demonstrated similar positive outcomes to those detailed in Section 4.2. Notably, its effectiveness becomes more pronounced as system complexity increases; in scenarios involving large classes, intricate functions, or multi-application interactions within microservices, \mymodel more reliably completes code editing tasks than alternative methods. We hypothesize that this superiority stems from its core design: by capturing the fine-grained user editing trajectory, \mymodel implicitly learns the developer's mental model and high-level understanding of the system's architecture—a rich signal that is absent in methods relying solely on static code snapshots.

\section{Related Work}

\paragraph{Code Edits with LLMs}                                 
Early transformer-based approaches extend code generation to editing tasks~\cite{Li2023InstructCoderIT,GRACE,Liu2024CoEdPilotRC,wei2024coeditor,Zhang2022OverwatchLP}, including AST-based methods~\cite{Chakraborty2018CODITCE}, retrieval-augmented frameworks~\cite{Liu2023AutomatedCE}, and versatile encoder-decoder models such
as CodeT5~\cite{Wang2021CodeT5IU}. Recent LLMs have made significant progress in code generation~\cite{Chen2021EvaluatingLL,Zeng2025ACECODERAC,Di2023CodeFuse13BAP,codefuse2025samplemattersleveragingmixtureofexperts,Jiang2024ROCODEIB,Tian2023FixingLL,moonshot_ai_2025}, code completion~\cite{liu2017neural,Alon2019StructuralLM,CodeBERT,Liu2020MultitaskLB,Chakraborty2021OnML,Lu2022ReACCAR,REPOFUSE,Naznin2023ANA,Wang2024LLMsML,Wan2024DoesYN,Wang2024RLCoderRL}, and other code-related tasks~\cite{liu2024automaticallyrecommendcodeupdates,Aleithan2024SWEBenchEC,Chen2025ACEBenchWW}, with Codex~\cite{Chen2021EvaluatingLL} demonstrating the ability to generate entire functions and applications from complex natural language prompts. However, these tools often lack optimization for code editing's strict requirements of low latency, high accuracy, and efficiency.


\paragraph{Specialized Code Edit} 
The advent of LLMs has spurred research in code editing, leading to the development of numerous approaches tailored specifically for code editing tasks~\cite{Li2023InstructCoderIT,GRACE,Liu2024CoEdPilotRC,wei2024coeditor,jin2025suggestingcodeeditsinteractive, wang2025reusegenerateacceleratingcode}.
InstructCoder~\cite{Li2023InstructCoderIT} introduces a dataset of 114,000 instruction-based triplets for LLM tuning, enhancing code editing accuracy for open-source models. However, it requires precise developer input, adding cognitive overhead.
CoEdPilot~\cite{Liu2024CoEdPilotRC} enhances code edit recommendations using historical edits and project context, but relies heavily on developer input and incurs high computational costs.
Unlike prompt-based models, \mymodel proactively predicts edit locations and content, enabling a seamless, instruction-free workflow.

\section{Conclusion}

We present \mymodel, an LLM framework for proactive, instruction-free code editing that learns from user editing trajectories. It achieves state-of-the-art performance (91.36\% Edit Similarity, 27.7\% Exact Match), serves 20,000+ developers at Ant Group with under 250ms latency and Acceptance Rates of 43.44\% and 51.55\%, and publicly releases its training datasets.

\clearpage

\bibliographystyle{ACM-Reference-Format}
\bibliography{reference}

@article{Di2023CodeFuse13BAP,
  title={CodeFuse-13B: A Pretrained Multi-Lingual Code Large Language Model},
  author={Peng Di and Jianguo Li and Hang Yu and Wei Jiang and Wenting Cai and Yang Cao and Chaoyu Chen and Dajun Chen and Hongwei Chen and Liang Chen and Gang Fan and Jie Gong and Zi Gong and Wen Hu and Tingting Guo and Zhichao Lei and Ting Li and Zheng Li and Ming Liang and Cong Liao and Bingchang Liu and Jiachen Liu and Zhiwei Liu and Shaojun Lu and Mingquan Shen and Guangpei Wang and Huan Wang and Zhi Yu Wang and Zhaogui Xu and Jiawei Yang and Qing Ye and Gehao Zhang and Yu Zhang and Zelin Zhao and Xunjin Zheng and Hailian Zhou and Lifu Zhu and Xianying Zhu},
  journal={2024 IEEE/ACM 46th International Conference on Software Engineering: Software Engineering in Practice (ICSE-SEIP)},
  year={2023},
  pages={418-429},
}

@article{Wang2024RLCoderRL,
  title={RLCoder: Reinforcement Learning for Repository-Level Code Completion},
  author={Yanlin Wang and Yanlin Wang and Daya Guo and Jiachi Chen and Ruikai Zhang and Yuchi Ma and Zibin Zheng},
  journal={2025 IEEE/ACM 47th International Conference on Software Engineering (ICSE)},
  year={2024},
  pages={1140-1152},
}

@article{Seed2025SeedCoderLT,
  title={Seed-Coder: Let the Code Model Curate Data for Itself},
  author={ByteDance Seed and Yuyu Zhang and Jing Su and Yifan Sun and Chenguang Xi and Xia Xiao and Shen Zheng and Anxiang Zhang and Kaibo Liu and Daoguang Zan and Tao Sun and Jinhua Zhu and Shulin Xin and Dong Huang and Yetao Bai and Lixin Dong and Chao Li and Jianchong Chen and Hanzhi Zhou and Yifan Huang and Guanghan Ning and Xierui Song and Jiaze Chen and Siyao Liu and Kai Shen and Liang Xiang and Yonghui Wu},
  journal={ArXiv},
  year={2025},
  volume={abs/2506.03524},
}

@article{Guo2024DeepSeekCoderWT,
  title={DeepSeek-Coder: When the Large Language Model Meets Programming - The Rise of Code Intelligence},
  author={Daya Guo and Qihao Zhu and Dejian Yang and Zhenda Xie and Kai Dong and Wentao Zhang and Guanting Chen and Xiao Bi and Yu Wu and Y. K. Li and Fuli Luo and Yingfei Xiong and Wenfeng Liang},
  journal={ArXiv},
  year={2024},
  volume={abs/2401.14196},
}

@article{Rozire2023CodeLO,
  title={Code Llama: Open Foundation Models for Code},
  author={Baptiste Rozi{\`e}re and Jonas Gehring and Fabian Gloeckle and Sten Sootla and Itai Gat and Xiaoqing Tan and Yossi Adi and Jingyu Liu and Tal Remez and J{\'e}r{\'e}my Rapin and Artyom Kozhevnikov and I. Evtimov and Joanna Bitton and Manish P Bhatt and Cris-tian Cant{\'o}n Ferrer and Aaron Grattafiori and Wenhan Xiong and Alexandre D'efossez and Jade Copet and Faisal Azhar and Hugo Touvron and Louis Martin and Nicolas Usunier and Thomas Scialom and Gabriel Synnaeve},
  journal={ArXiv},
  year={2023},
  volume={abs/2308.12950},
}

@article{Ding2024VulnerabilityDW,
  title={Vulnerability Detection with Code Language Models: How Far are We?},
  author={Yangruibo Ding and Yanjun Fu and Omniyyah Ibrahim and Chawin Sitawarin and Xinyun Chen and Basel Alomair and David A. Wagner and Baishakhi Ray and Yizheng Chen},
  journal={2025 IEEE/ACM 47th International Conference on Software Engineering (ICSE)},
  year={2024},
  pages={1729-1741},
}

@article{REPOFUSE,
  title={REPOFUSE: Repository-Level Code Completion with Fused Dual Context},
  author={Ming Liang and Xiaoheng Xie and Gehao Zhang and Xunjin Zheng and Peng Di and Wei Jiang and Hongwei Chen and Chengpeng Wang and Gang Fan},
  journal={ArXiv},
  year={2024},
  volume={abs/2402.14323},
}

@INPROCEEDINGS{6693078,
  author={Nguyen, Hoan Anh and Nguyen, Anh Tuan and Nguyen, Tung Thanh and Nguyen, Tien N. and Rajan, Hridesh},
  booktitle={2013 28th IEEE/ACM International Conference on Automated Software Engineering (ASE)}, 
  title={A study of repetitiveness of code changes in software evolution}, 
  year={2013},
  volume={},
  number={},
  pages={180-190},
  doi={10.1109/ASE.2013.6693078}}

@article{Zhang2022CoditT5PF,
  title={CoditT5: Pretraining for Source Code and Natural Language Editing},
  author={Jiyang Zhang and Sheena Panthaplackel and Pengyu Nie and Junyi Jessy Li and Milos Gligoric},
  journal={Proceedings of the 37th IEEE/ACM International Conference on Automated Software Engineering},
  year={2022},
}

@misc{zheng2024sglangefficientexecutionstructured,
      title={SGLang: Efficient Execution of Structured Language Model Programs}, 
      author={Lianmin Zheng and Liangsheng Yin and Zhiqiang Xie and Chuyue Sun and Jeff Huang and Cody Hao Yu and Shiyi Cao and Christos Kozyrakis and Ion Stoica and Joseph E. Gonzalez and Clark Barrett and Ying Sheng},
      year={2024},
      eprint={2312.07104},
      archivePrefix={arXiv},
      primaryClass={cs.AI},
}

@article{Chen2023AcceleratingLL,
  title={Accelerating Large Language Model Decoding with Speculative Sampling},
  author={Charlie Chen and Sebastian Borgeaud and Geoffrey Irving and Jean-Baptiste Lespiau and L. Sifre and John M. Jumper},
  journal={ArXiv},
  year={2023},
  volume={abs/2302.01318},
}

@inproceedings{Speculative,
  title={Fast Inference from Transformers via Speculative Decoding},
  author={Yaniv Leviathan and Matan Kalman and Yossi Matias},
  booktitle={International Conference on Machine Learning},
  year={2022},
}

@misc{
liu2017neural,
title={Neural Code Completion},
author={Chang Liu and Xin Wang and Richard Shin and Joseph E. Gonzalez and Dawn Song},
year={2017},
}

@article{Alon2019StructuralLM,
  title={Structural Language Models for Any-Code Generation},
  author={Uri Alon and Roy Sadaka and Omer Levy and Eran Yahav},
  journal={ArXiv},
  year={2019},
  volume={abs/1910.00577},
}

@article{Liu2020MultitaskLB,
  title={Multi-task Learning based Pre-trained Language Model for Code Completion},
  author={F. Liu and Ge Li and Yunfei Zhao and Zhi Jin},
  journal={2020 35th IEEE/ACM International Conference on Automated Software Engineering (ASE)},
  year={2020},
  pages={473-485},
}

@article{CodeBERT,
  title={CodeBERT: A Pre-Trained Model for Programming and Natural Languages},
  author={Zhangyin Feng and Daya Guo and Duyu Tang and Nan Duan and Xiaocheng Feng and Ming Gong and Linjun Shou and Bing Qin and Ting Liu and Daxin Jiang and Ming Zhou},
  journal={ArXiv},
  year={2020},
  volume={abs/2002.08155},
}

@article{Liu2023AutomatedCE,
  title={Automated Code Editing With Search-Generate-Modify},
  author={Changshuo Liu and Pelin Çetin and Yogesh Patodia and Saikat Chakraborty and Yangruibo Ding and Baishakhi Ray},
  journal={IEEE Transactions on Software Engineering},
  year={2023},
  volume={50},
  pages={1675-1686},
}

@article{Chakraborty2018CODITCE,
  title={CODIT: Code Editing With Tree-Based Neural Models},
  author={Saikat Chakraborty and Yangruibo Ding and Miltiadis Allamanis and Baishakhi Ray},
  journal={IEEE Transactions on Software Engineering},
  year={2018},
  volume={48},
  pages={1385-1399},
}

@article{Wang2021CodeT5IU,
  title={CodeT5: Identifier-aware Unified Pre-trained Encoder-Decoder Models for Code Understanding and Generation},
  author={Yue Wang and Weishi Wang and Shafiq R. Joty and Steven C. H. Hoi},
  journal={ArXiv},
  year={2021},
  volume={abs/2109.00859},
}

@misc{codefuse2025samplemattersleveragingmixtureofexperts,
      title={Every Sample Matters: Leveraging Mixture-of-Experts and High-Quality Data for Efficient and Accurate Code LLM}, 
      author={Codefuse and Ling Team  and Wenting Cai and Yuchen Cao and Chaoyu Chen and Chen Chen and Siba Chen and Qing Cui and Peng Di and Junpeng Fang and Zi Gong and Ting Guo and Zhengyu He and Yang Huang and Cong Li and Jianguo Li and Zheng Li and Shijie Lian and BingChang Liu and Songshan Luo and Shuo Mao and Min Shen and Jian Wu and Jiaolong Yang and Wenjie Yang and Tong Ye and Hang Yu and Wei Zhang and Zhenduo Zhang and Hailin Zhao and Xunjin Zheng and Jun Zhou},
      year={2025},
      eprint={2503.17793},
      archivePrefix={arXiv},
      primaryClass={cs.LG},
}

@inproceedings{Li2023InstructCoderIT,
  title={InstructCoder: Instruction Tuning Large Language Models for Code Editing},
  author={Kaixin Li and Qisheng Hu and Xu Zhao and Hui Chen and Yuxi Xie and Tiedong Liu and Qizhe Xie and Junxian He},
  booktitle={Annual Meeting of the Association for Computational Linguistics},
  year={2023},
}

@inproceedings{GRACE,
author = {Gupta, Priyanshu and Khare, Avishree and Bajpai, Yasharth and Chakraborty, Saikat and Gulwani, Sumit and Kanade, Aditya and Radhakrishna, Arjun and Soares, Gustavo and Tiwari, Ashish},
title = {Grace: Language Models Meet Code Edits},
year = {2023},
isbn = {9798400703270},
publisher = {Association for Computing Machinery},
address = {New York, NY, USA},
doi = {10.1145/3611643.3616253},
booktitle = {Proceedings of the 31st ACM Joint European Software Engineering Conference and Symposium on the Foundations of Software Engineering},
pages = {1483–1495},
numpages = {13},
location = {San Francisco, CA, USA},
series = {ESEC/FSE 2023}
}

@article{Liu2024CoEdPilotRC,
  title={CoEdPilot: Recommending Code Edits with Learned Prior Edit Relevance, Project-wise Awareness, and Interactive Nature},
  author={Chenyan Liu and Yufan Cai and Yun Lin and Yuhuan Huang and Yunrui Pei and Bo Jiang and Ping Yang and Jin Song Dong and Hong Mei},
  journal={Proceedings of the 33rd ACM SIGSOFT International Symposium on Software Testing and Analysis},
  year={2024},
}

@inproceedings{
wei2024coeditor,
title={Coeditor: Leveraging Repo-level Diffs for Code Auto-editing},
author={Jiayi Wei and Greg Durrett and Isil Dillig},
booktitle={The Twelfth International Conference on Learning Representations},
year={2024},
}

@misc{kimiteam2025kimik15scalingreinforcement,
      title={Kimi k1.5: Scaling Reinforcement Learning with LLMs}, 
      author={Kimi Team},
      year={2025},
      eprint={2501.12599},
      archivePrefix={arXiv},
      primaryClass={cs.AI},
}

@article{Chakraborty2021OnML,
  title={On Multi-Modal Learning of Editing Source Code},
  author={Saikat Chakraborty and Baishakhi Ray},
  journal={2021 36th IEEE/ACM International Conference on Automated Software Engineering (ASE)},
  year={2021},
  pages={443-455},
}

@article{InstructGPT,
  title={Training language models to follow instructions with human feedback},
  author={Long Ouyang and Jeff Wu and Xu Jiang and Diogo Almeida and Carroll L. Wainwright and Pamela Mishkin and Chong Zhang and Sandhini Agarwal and Katarina Slama and Alex Ray and John Schulman and Jacob Hilton and Fraser Kelton and Luke E. Miller and Maddie Simens and Amanda Askell and Peter Welinder and Paul Francis Christiano and Jan Leike and Ryan J. Lowe},
  journal={ArXiv},
  year={2022},
  volume={abs/2203.02155},
}

@article{Chen2021EvaluatingLL,
  title={Evaluating Large Language Models Trained on Code},
  author={Mark Chen and Jerry Tworek and Heewoo Jun and Qiming Yuan and Henrique Pond{\'e} and Jared Kaplan and Harrison Edwards and Yura Burda and Nicholas Joseph and Greg Brockman and Alex Ray and Raul Puri and Gretchen Krueger and Michael Petrov and Heidy Khlaaf and Girish Sastry and Pamela Mishkin and Brooke Chan and Scott Gray and Nick Ryder and Mikhail Pavlov and Alethea Power and Lukasz Kaiser and Mo Bavarian and Clemens Winter and Phil Tillet and Felipe Petroski Such and David W. Cummings and Matthias Plappert and Fotios Chantzis and Elizabeth Barnes and Ariel Herbert-Voss and William H. Guss and Alex Nichol and Igor Babuschkin and Suchir Balaji and Shantanu Jain and Andrew Carr and Jan Leike and Josh Achiam and Vedant Misra and Evan Morikawa and Alec Radford and Matthew M. Knight and Miles Brundage and Mira Murati and Katie Mayer and Peter Welinder and Bob McGrew and Dario Amodei and Sam McCandlish and Ilya Sutskever and Wojciech Zaremba},
  journal={ArXiv},
  year={2021},
  volume={abs/2107.03374},
}

@misc{moonshot_ai_2025,
	author       = { Moonshot AI },
	title        = { Kimi-K2-Instruct (Revision 2f7e011) },
	year         = 2025,
	url          = { https://huggingface.co/moonshotai/Kimi-K2-Instruct },
	doi          = { 10.57967/hf/5976 },
	publisher    = { Hugging Face }
}

@article{Aleithan2024SWEBenchEC,
  title={SWE-Bench+: Enhanced Coding Benchmark for LLMs},
  author={Reem Aleithan and Haoran Xue and Mohammad Mahdi Mohajer and Elijah Nnorom and Gias Uddin and Song Wang},
  journal={ArXiv},
  year={2024},
  volume={abs/2410.06992},
}

@inproceedings{Chen2025ACEBenchWW,
  title={ACEBench: Who Wins the Match Point in Tool Usage?}, 
      author={Chen Chen and Xinlong Hao and Weiwen Liu and Xu Huang and Xingshan Zeng and Shuai Yu and Dexun Li and Shuai Wang and Weinan Gan and Yuefeng Huang and Wulong Liu and Xinzhi Wang and Defu Lian and Baoqun Yin and Yasheng Wang and Wu Liu},
      year={2025},
      eprint={2501.12851},
      archivePrefix={arXiv},
      primaryClass={cs.CL},
}

@article{Zeng2025ACECODERAC,
  title={ACECODER: Acing Coder RL via Automated Test-Case Synthesis},
  author={Huaye Zeng and Dongfu Jiang and Haozhe Wang and Ping Nie and Xiaotong Chen and Wenhu Chen},
  journal={ArXiv},
  year={2025},
  volume={abs/2502.01718},
}

@article{Jiang2024ROCODEIB,
  title={ROCODE: Integrating Backtracking Mechanism and Program Analysis in Large Language Models for Code Generation},
  author={Xue Jiang and Yihong Dong and Yongding Tao and Huanyu Liu and Zhi Jin and Wenpin Jiao and Ge Li},
  journal={ArXiv},
  year={2024},
  volume={abs/2411.07112},
}

@inproceedings{Tian2023FixingLL,
  title={Fixing Large Language Models' Specification Misunderstanding for Better Code Generation},
  author={Zhao Tian and Junjie Chen},
  year={2023},
}

@article{Wang2024LLMsML,
  title={LLMs Meet Library Evolution: Evaluating Deprecated API Usage in LLM-Based Code Completion},
  author={Chong Wang and Kaifeng Huang and Jian Zhang and Yebo Feng and Lyuye Zhang and Yang Liu and Xin Peng},
  journal={2025 IEEE/ACM 47th International Conference on Software Engineering (ICSE)},
  year={2024},
  pages={885-897},
}

@article{Lu2022ReACCAR,
  title={ReACC: A Retrieval-Augmented Code Completion Framework},
  author={Shuai Lu and Nan Duan and Hojae Han and Daya Guo and Seung-won Hwang and Alexey Svyatkovskiy},
  journal={ArXiv},
  year={2022},
  volume={abs/2203.07722},
}

@article{Naznin2023ANA,
  title={A Naive Approach for Automatic Line-level Code Completion},
  author={Shamima Naznin and Manishankar Mondal},
  journal={2023 IEEE 9th International Women in Engineering (WIE) Conference on Electrical and Computer Engineering (WIECON-ECE)},
  year={2023},
  pages={137-142},
}

@article{Wan2024DoesYN,
  title={Does Your Neural Code Completion Model Use My Code? A Membership Inference Approach},
  author={Yao Wan and Guanghua Wan and Shijie Zhang and Hongyu Zhang and Yulei Sui and Pan Zhou and Hai Jin and Lichao Sun},
  journal={ACM Transactions on Software Engineering and Methodology},
  year={2024},
}

@misc{qwen3technicalreport,
      title={Qwen3 Technical Report}, 
      author={Qwen Team},
      year={2025},
      eprint={2505.09388},
      archivePrefix={arXiv},
      primaryClass={cs.CL},
      url={https://arxiv.org/abs/2505.09388}, 
}

@article{Yang2024Qwen25TR,
  title={Qwen2.5 Technical Report},
  author={Qwen Team},
  journal={ArXiv},
  year={2024},
  volume={abs/2412.15115},
}

@inproceedings{TheC3,
  title={The Claude 3 Model Family: Opus, Sonnet, Haiku},
  author={Anthropic},
  year={2024},
  url={https://api.semanticscholar.org/CorpusID:268232499}
}

@misc{guo2025codeeditorbenchevaluatingcodeediting,
      title={CodeEditorBench: Evaluating Code Editing Capability of Large Language Models}, 
      author={Jiawei Guo and Ziming Li and Xueling Liu and Kaijing Ma and Tianyu Zheng and Zhouliang Yu and Ding Pan and Yizhi LI and Ruibo Liu and Yue Wang and Shuyue Guo and Xingwei Qu and Xiang Yue and Ge Zhang and Wenhu Chen and Jie Fu},
      year={2025},
      eprint={2404.03543},
      archivePrefix={arXiv},
      primaryClass={cs.SE},
}

@misc{copilot,
      title={GitHub Copilot}, 
      author={Copilot},
      year={2023},
      url={https://github.com/features/copilot}, 
}

@misc{cursor,
      title={Cursor}, 
      author={Cursor},
      year={2024},
      url={https://cursor.com}, 
}

@misc{zeta,
      title={Zeta}, 
      author={Zed},
      year={2025},
      url={https://huggingface.co/zed-industries/zeta}, 
}

@book{card2018psychology,
  title={The psychology of human-computer interaction},
  author={Card, Stuart K},
  year={2018},
  publisher={Crc Press}
}

@article{Zhang2022OverwatchLP,
  title={Overwatch: learning patterns in code edit sequences},
  author={Yuhao Zhang and Yasharth Bajpai and Priyanshu Gupta and Ameya Ketkar and Miltiadis Allamanis and Titus Barik and Sumit Gulwani and Arjun Radhakrishna and Mohammad Raza and Gustavo Soares and Ashish Tiwari},
  journal={Proceedings of the ACM on Programming Languages},
  year={2022},
  volume={6},
  pages={395 - 423},
}

@misc{jin2025suggestingcodeeditsinteractive,
      title={Suggesting Code Edits in Interactive Machine Learning Notebooks Using Large Language Models}, 
      author={Bihui Jin and Jiayue Wang and Pengyu Nie},
      year={2025},
      eprint={2501.09745},
      archivePrefix={arXiv},
      primaryClass={cs.SE},
}

@misc{liu2024automaticallyrecommendcodeupdates,
      title={Automatically Recommend Code Updates: Are We There Yet?}, 
      author={Yue Liu and Chakkrit Tantithamthavorn and Yonghui Liu and Patanamon Thongtanunam and Li Li},
      year={2024},
      eprint={2209.07048},
      archivePrefix={arXiv},
      primaryClass={cs.SE},
}

@misc{wang2025reusegenerateacceleratingcode,
      title={Reuse or Generate? Accelerating Code Editing via Edit-Oriented Speculative Decoding}, 
      author={Peiding Wang and Li Zhang and Fang Liu and Yinghao Zhu and Wang Xu and Lin Shi and Xiaoli Lian and Minxiao Li and Bo Shen and An Fu},
      year={2025},
      eprint={2506.02780},
      archivePrefix={arXiv},
      primaryClass={cs.SE},
}

@misc{yu2025dapoopensourcellmreinforcement,
      title={DAPO: An Open-Source LLM Reinforcement Learning System at Scale}, 
      author={Qiying Yu and Zheng Zhang and Ruofei Zhu and Yufeng Yuan and Xiaochen Zuo and Yu Yue and Weinan Dai and Tiantian Fan and Gaohong Liu and Lingjun Liu and Xin Liu and Haibin Lin and Zhiqi Lin and Bole Ma and Guangming Sheng and Yuxuan Tong and Chi Zhang and Mofan Zhang and Wang Zhang and Hang Zhu and Jinhua Zhu and Jiaze Chen and Jiangjie Chen and Chengyi Wang and Hongli Yu and Yuxuan Song and Xiangpeng Wei and Hao Zhou and Jingjing Liu and Wei-Ying Ma and Ya-Qin Zhang and Lin Yan and Mu Qiao and Yonghui Wu and Mingxuan Wang},
      year={2025},
      eprint={2503.14476},
      archivePrefix={arXiv},
      primaryClass={cs.LG},
}

@inproceedings{
wei2022finetuned,
title={Finetuned Language Models are Zero-Shot Learners},
author={Jason Wei and Maarten Bosma and Vincent Zhao and Kelvin Guu and Adams Wei Yu and Brian Lester and Nan Du and Andrew M. Dai and Quoc V Le},
booktitle={International Conference on Learning Representations},
year={2022},
}

@inproceedings{ren2025peaceefficientprojectlevelefficiency,
      title={PEACE: Towards Efficient Project-Level Efficiency Optimization via Hybrid Code Editing}, 
      author={Xiaoxue Ren and Jun Wan and Yun Peng and Zhongxin Liu and Ming Liang and Dajun Chen and Wei Jiang and Yong Li},
      year={2025},
      journal={the IEEE/ACM International Conference on Automated Software Engineering (ASE)},
}

@inproceedings{Song_2024,
   title={Revisiting Code Similarity Evaluation with Abstract Syntax Tree Edit Distance},
   url={http://dx.doi.org/10.18653/v1/2024.acl-short.3},
   DOI={10.18653/v1/2024.acl-short.3},
   booktitle={Proceedings of the 62nd Annual Meeting of the Association for Computational Linguistics (Volume 2: Short Papers)},
   publisher={Association for Computational Linguistics},
   author={Song, Yewei and Lothritz, Cedric and Tang, Xunzhu and Bissyandé, Tegawendé and Klein, Jacques},
   year={2024},
   pages={38–46} }

@misc{tao2025codegraphmodelcgm,
      title={Code Graph Model (CGM): A Graph-Integrated Large Language Model for Repository-Level Software Engineering Tasks}, 
      author={Hongyuan Tao and Ying Zhang and Zhenhao Tang and Hongen Peng and Xukun Zhu and Bingchang Liu and Yingguang Yang and Ziyin Zhang and Zhaogui Xu and Haipeng Zhang and Linchao Zhu and Rui Wang and Hang Yu and Jianguo Li and Peng Di},
      year={2025},
      eprint={2505.16901},
      archivePrefix={arXiv},
      primaryClass={cs.SE},
      url={https://arxiv.org/abs/2505.16901}, 
}

@inproceedings{zhang-etal-2025-galla,
    title = "{GALL}a: Graph Aligned Large Language Models for Improved Source Code Understanding",
    author = "Zhang, Ziyin  and
      Yu, Hang  and
      Lee, Sage  and
      Di, Peng  and
      Li, Jianguo  and
      Wang, Rui",
    booktitle = "Proceedings of the 63rd Annual Meeting of the Association for Computational Linguistics (Volume 1: Long Papers)",
    month = jul,
    year = "2025",
    address = "Vienna, Austria",
    publisher = "Association for Computational Linguistics",
    url = "https://aclanthology.org/2025.acl-long.676/",
    doi = "10.18653/v1/2025.acl-long.676",
    pages = "13784--13802",
    ISBN = "979-8-89176-251-0",
}

@inproceedings{10.1145/3691620.3694999,
author = {Li, Cong and Xu, Zhaogui and Di, Peng and Wang, Dongxia and Li, Zheng and Zheng, Qian},
title = {Understanding Code Changes Practically with Small-Scale Language Models},
year = {2024},
isbn = {9798400712487},
publisher = {Association for Computing Machinery},
address = {New York, NY, USA},
url = {https://doi.org/10.1145/3691620.3694999},
doi = {10.1145/3691620.3694999},
booktitle = {Proceedings of the 39th IEEE/ACM International Conference on Automated Software Engineering},
pages = {216–228},
numpages = {13},
location = {Sacramento, CA, USA},
series = {ASE '24}
}


\end{document}